\documentclass[sigconf,natbib=true]{acmart} 

\AtBeginDocument{%
  }

\setcopyright{acmlicensed}
\copyrightyear{2024}
\acmYear{2024}
\acmDOI{XXXXXXX.XXXXXXX}

\acmConference[COMPASS '24]{ACM SIGCAS/SIGCHI Conference on Computing and Sustainable Societies}{July 8 - July 11, 2024}{New Delhi, India}
\acmISBN{978-1-4503-XXXX-X/18/06}

\usepackage{soul}
\usepackage{enumitem}

\begin{document}

\title{Comuniqa : Exploring Large Language Models for Improving English Speaking Skills}

\author{Manas Mhasakar}
\authornote{Both authors contributed equally to this research.}
\email{t-mmhasakar@microsoft.com}
\orcid{0009-0009-8609-5189}
\affiliation{%
  \institution{Microsoft Research}
  \city{Bengaluru}
  \country{India}
}

\author{Shikhar Sharma}
\authornotemark[1]
\email{shikhar20121@iiitd.ac.in}
\affiliation{%
  \institution{IIIT Delhi}
  \city{New Delhi}
  \country{India}
  }

\author{Apurv Mehra}
\email{apurv@blendnet.ai}
\affiliation{%
  \institution{BlendNet AI}
  \city{Bengaluru}
  \country{India}}

\author{Utkarsh Venaik}
\email{utkarsh21570@iiitd.ac.in}
\affiliation{%
  \institution{IIIT Delhi}
  \city{New Delhi}
  \country{India}
  }

\author{Ujjwal Singhal}
\email{ujjwal21434@iiitd.ac.in}
\affiliation{%
  \institution{IIIT Delhi}
  \city{New Delhi}
  \country{India}
  }

\author{Dhruv Kumar}
\email{dhruv.kumar@iiitd.ac.in}
\affiliation{%
  \institution{IIIT Delhi}
  \city{New Delhi}
  \country{India}
  }

\author{Kashish Mittal}
\email{kmittal@microsoft.com}
\affiliation{%
  \institution{Microsoft Research}
  \city{Bengaluru}
  \country{India}
}
\renewcommand{\shortauthors}{
  Mhasakar, Sharma, et al.
}

\begin{abstract}
In this paper, we investigate the potential of Large Language Models (LLMs) to improve English speaking skills. This is particularly relevant in countries like India, where English is crucial for academic, professional, and personal communication but remains a non-native language for many. Traditional methods for enhancing speaking skills often rely on human experts, which can be limited in terms of scalability, accessibility, and affordability. Recent advancements in Artificial Intelligence (AI) offer promising solutions to overcome these limitations.


We propose Comuniqa, a novel LLM-based system designed to enhance English speaking skills. We adopt a human-centric evaluation approach, comparing Comuniqa with the feedback and instructions provided by human experts.
In our evaluation, we divide the participants in three groups: those who use LLM-based system for improving speaking skills, those guided by human experts for the same task and those who utilize both the LLM-based system as well as the human experts. Using surveys, interviews, and actual study sessions, we provide a detailed perspective on the effectiveness of different learning modalities. Our preliminary findings suggest that while LLM-based systems have commendable accuracy, they lack human-level cognitive capabilities, both in terms of accuracy and empathy. 
Nevertheless, Comuniqa represents a significant step towards achieving \emph{Sustainable Development Goal 4: Quality Education} by providing a valuable learning tool for individuals who may not have access to human experts for improving their speaking skills.


\end{abstract}

\ccsdesc[300]{Social and professional topics~Sustainability}
\ccsdesc[500]{Human-centered computing~Empirical studies in HCI}

\keywords{Large Language Models, Speaking Skills, Human-LLM Collaboration, Speech Interface}

\received{20 February 2007}
\received[revised]{12 March 2009}
\received[accepted]{5 June 2009}

\maketitle

\section{Introduction}\label{sec:intro}
Speaking skills are paramount to effective interaction in the multifaceted landscape of interpersonal communication. They serve as the cornerstone for successful collaboration, persuasion, and the conveyance of ideas. The ability to express thoughts eloquently and coherently is integral in academic, professional, and personal spheres.

Some of the conventional methods for honing speaking skills include (1) going through formal training through public speaking courses and workshops, (2) taking constructive feedback and engaging in conversations with friends, colleagues and experts, (3) recording and reviewing one's speech, and (4) enhancing vocabulary and language expressions by regular reading and listening to a variety of sources \cite{goodreadsPublicSpeaking, tuhovsky2015communication, springerEffectsReciprocal}.

The aforementioned methodologies provide valuable insights and practical experiences; however, they are not without limitations. This challenge is particularly salient in a country such as India, where English is not the native language for the majority of the population. The dearth of opportunities to engage in spoken English practice may detrimentally affect individuals' employment prospects, educational opportunities, social interactions etc. Experts and proficient speakers are not easily accessible and affordable, and individuals may harbor apprehensions about potential judgment. Self-assessment through recording, while a valuable practice, may prove inadequate without the inclusion of external feedback from experts. Conventional speaking courses and workshops, in certain instances, may not offer the level of personalized feedback essential for individualized skill enhancement. Moreover, the efficacy of these conventional approaches can be constrained by the inability to address diverse individual learning needs. 

Advancements in artificial intelligence present a transformative potential in revolutionizing the enhancement of speaking skills. AI systems can offer scalable, accessible, and personalized learning experiences. These models, equipped with vast linguistic knowledge, can analyze spoken language nuances, provide constructive feedback and adapt to individual learning styles, offering a dynamic and tailored approach to enhancing speaking skills. Several user studies have been conducted to evaluate AI's benefits in improving speaking skills \cite{ruan2021englishbot, wang2020voicecoach, tanveer2015rhema, junaidi2020artificial, Trinh2017robocop}. 




In this work, we explore the viability of utilizing the recent advances in Large Language Models (LLMs) \cite{brown2020language, wei2022emergent, bubeck2023sparks} to improve English speaking skills. LLMs are being evaluated in various contexts, but there is limited work on evaluating their potential for improving English speaking skills. We aim to bridge this gap by taking a human-centric approach to evaluate the value addition provided by LLM-based applications for improving English speaking skills. We also compare the strengths and weaknesses of these LLM-based applications with those of human experts in English speaking skills.

Our methodology uses purposive and random sampling for participant recruitment, categorizing them into three distinct groups: those who use LLM-enabled sysstem for improving English speaking skills (Group 1), those guided by human experts for the same task (Group 2), and those who utilize both the LLM-enabled system as well as the human experts (Group 3). We conduct pre-study and post-study surveys, interviews, and actual study sessions for each participant, facilitating a comprehensive understanding of their experiences. The data collected is then analyzed both qualitatively and quantitatively, providing a detailed perspective on the effectiveness of different learning modalities.

This study is guided by the following research questions in the context of improving English speaking skills:
\begin{itemize}
    \item \textbf{RQ1:} What are the strengths and weaknesses of LLM-based system and human experts as perceived by the participants?
    \item \textbf{RQ2:} How can a LLM-based system aid the role played by human experts in speaking practise/oral communication?
\end{itemize}



 The key contributions of our work include: \textbf{(1)} a novel platform for improving English speaking skills designed using the latest advances in LLMs, \textbf{(2)} a novel user study evaluating users' preferences for LLM-based systems and human experts, and \textbf{(3)} understanding users' preferences and the effectiveness of LLM and expert-based systems, thus providing design considerations for such systems.
 This paper presents initial findings of our work-in-progress. Findings derived from a research study comprising 34 participants suggest that, although LLM-based systems exhibit commendable accuracy and depth, they fall short of matching the human cognitive capabilities, both in accuracy as well as empathy. Nevertheless, this paper takes a step towards achieving \emph{Sustainable Development Goal 4: Quality Education} by providing a valuable learning tool for individuals who may not have access to human experts for improving their English speaking skills.
\section{Related Work}\label{sec:related_work}




\subsection{Human-Centric Studies on AI for improving speaking skills}
Various user studies have focused on evaluating the benefits of AI-based techniques in enhancing speaking skills in different settings \cite{ruan2021englishbot, wang2020voicecoach, tanveer2015rhema, junaidi2020artificial, Trinh2017robocop}. Ruan et al \cite{ruan2021englishbot} and Junaidi et al \cite{junaidi2020artificial} propose AI apps for improving English speaking for non-native English speakers. Other studies \cite{wang2020voicecoach, tanveer2015rhema, Trinh2017robocop} focus on providing feedback in very specific scenarios such as live delivery of public speech or presentation rehersals. These studies focus on providing feedback to the user on various aspects of speech such as modulation, speaking rate, pitch and volume. 
All of these studies consider two groups of participants: one group which is exposed only to the AI-based app and the other group which is not exposed to AI-based app (may be exposed to human experts or may ask the participants for self-practice). In contrast, our study not only considers the two groups of participants but also evaluates a third group having participants which are exposed to both AI-based app and human experts. This additional group helps us to evaluate the possibility of combining feedback from both AI and human experts to enhance overall learning outcomes in the context of oral communication skills. Finally, we also aim to evaluate the specific advantages of the recent advances in LLMs which have not been explored by the existing studies.

\subsection{Human-Centric Studies on LLMs}
Existing research has explored humans using or collaborating with LLMs in a variety of contexts. Shen et al \cite{shen2023parachute}, Jakesch et al \cite{Jakesch2023}, Chung et al \cite{Chung2022}, 
and Yuan et al \cite{yuan2022wordcraft} evaluate humans collaborating with LLMs for different types of writing tasks. Becker et al \cite{becker2023programming} discuss how students can collaborate with code generation models to learn programming. Li et al \cite{li2023collaborative} investigates the synergy between humans and LLMs in evaluating open-ended natural language generation tasks. The paper uses a checklist of task-specific criteria for detailed text evaluation wherein LLMs do the initial evaluation followed by human scrutiny to improve the overall evaluation process. Petridis et al \cite{Petridis2023} propose an LLM-based conversational tool to help journalists in exploring various news angles from different types of documents. Their tool substantially reduced the cognitive load of the participants as compared to prior tools. Ashby et al \cite{Ashby2023} discuss collaboration between humans and AI systems in the context of gaming. Ruan et al \cite{Ruan2019}  developed a dialogue-based conversational agent resulting in increased learning, suggesting that educational chatbot systems may have beneficial use for learning outside of traditional settings. Valencia et al \cite{Valencia2023} found that using LLMs can reduce time and effort for users of augmentative and alternative communication (AAC) devices. Peng et al \cite{peng2023storyfier} propose a tool that leverages text generation models to support vocabulary learning. In contrast to the above mentioned studies, our work explores the effectiveness of utilizing LLMs for enhancing oral communication skills of any user. Additionally, we explore if employing subject matter experts alongisde LLMs further lead to an improved learning experience.





\section{System}\label{sec:system}
To explore the role of LLMs in enhancing speaking skills, we developed a mobile application called Comuniqa \cite{googleComuniqaApps}. The app enables users to practice speaking English and receive instant feedback. The app can be downloaded by anyone from the Google Play Store \cite{googleAndroidApps}. The mobile app leverages LLMs along with other state-of-the-art NLP (Natural Language Processing) and speech models for analyzing user speech and providing feedback to them. The overall score of the user is modelled on the lines of IELTS evaluation scheme \cite{IELTSSpeaking}.
This section provides a brief overview of our application’s architecture as well as the methodology for analyzing and evaluating user speech.

\subsection{Application Architecture}
The application's logic is implemented in Python, with the backend developed using the Django framework. We utilized React Native, a framework for developing the application front-end. We followed the principles of user accessibility during front-end development for better user experience. 
We evaluate pronunciation, pace and pitch directly from the user audio while we also convert the recorded user speech to text for evaluating other metrics such as Emotion, Sentiment, Confidence, Grammar, Coherence, Filler words, Fluency.

\noindent\textbf{Speech-To-Text Conversion:} For this, we employed the Whisper Timestamped model which is based on OpenAI's Whisper Model \cite{radford2023whisper, lintoai2023whispertimestamped, JSSv031i07}. This model accurately transcribes the speech audio along with timestamps. 
To identify and mark awkward pauses, we wrote an algorithm which calculates user pauses during speech and replaces awkward pause intervals with a unique character for user-friendly display. 

\subsection{LLM-Based Analysis}
The LLM-based analysis involves assessing user speech text (converted from user audio input) across various metrics encompassing vocabulary, fluency, sentiment, pronunciation, coherence, grammar, and overall summary. Some of these metrics are explained below:

\noindent\textbf{Vocabulary:} We developed modules for vocabulary analysis using public datasets for fetching CEFR level \cite{cefrWordLevel}  of each individual word in the users transcript and an open source pretrained model \cite{cefrPredictor} for predicting the overall CEFR score of the response. 

\noindent\textbf{Grammar:} For grammar evaluation, the user transcript undergoes scrutiny through few-shot learning \cite{brown2020language}, prompting the LLM-model with 20 predefined error categories accompanied by examples. The LLM then generates a score along with both incorrect and corrected phrases. 

\noindent\textbf{Coherence:} This involves prompts based on the completeness, relevance, and logical flow of the user's response, yielding an overall score and boolean scores for each parameter, accompanied by reasoning. 

\noindent\textbf{Filler words:} Evaluating filler words incorporates few-shot learning with 15 example pairs and a list of common fillers \cite{MicrosoftFillerWords}. The LLM outputs an overall score, total filler word count, and modified output with delimiters indicating filler word usage for the user over the UI. 

\noindent\textbf{Fluency:} Fluency, determined by pitch, pronunciation, pace, awkward pauses, and filler words, is consolidated into an overall fluency score.



\subsection{Analysis based on other AI models}
This section discusses the metrics which are evaluated using the state-of-the-art NLP and speech models (i.e. without any LLM).
\noindent\textbf{Pace and Pitch:} The application analyzes and visualizes user's input pace and pitch. The Whisper-Timestamped model's outputs facilitate precise calculations via the timestamps which are then used to calculate overall moving average of the user pace. This is then presented graphically in our mobile app's user interface, offering users a detailed understanding of their speech dynamics. To calculate the pitch of the user, the Praat-Parselmouth library \cite{praat, parselmouth} was used for analysing Pitch and return an overall score and a graph similar to pace of the user. 

\noindent\textbf{Pronunciation:} For evaluating pronunciation, Microsoft Azure’s AI Cognitive Services \cite{azureSpeech} were utilized which outputs the correct pronunciation of a word which the user has pronounced incorrectly. As a part of the UI, the user gets to listen to their original and correct pronunciation.


\begin{figure*}[ht]
    \includegraphics[width=0.6\textwidth, height=0.2\textheight]{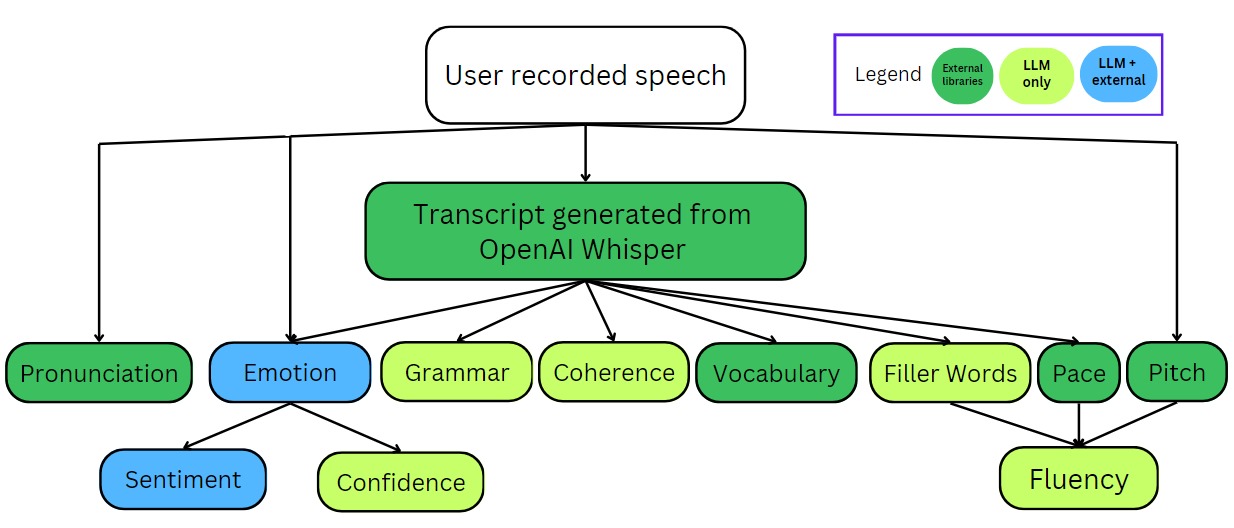}
    \caption{Block level diagram of the system infrastructure}
    \label{fig:sample}
\end{figure*}

\subsection{User Workflow}
\begin{figure*}
\centering
\begin{minipage}{.21\textwidth}
  \includegraphics[width=1\linewidth]{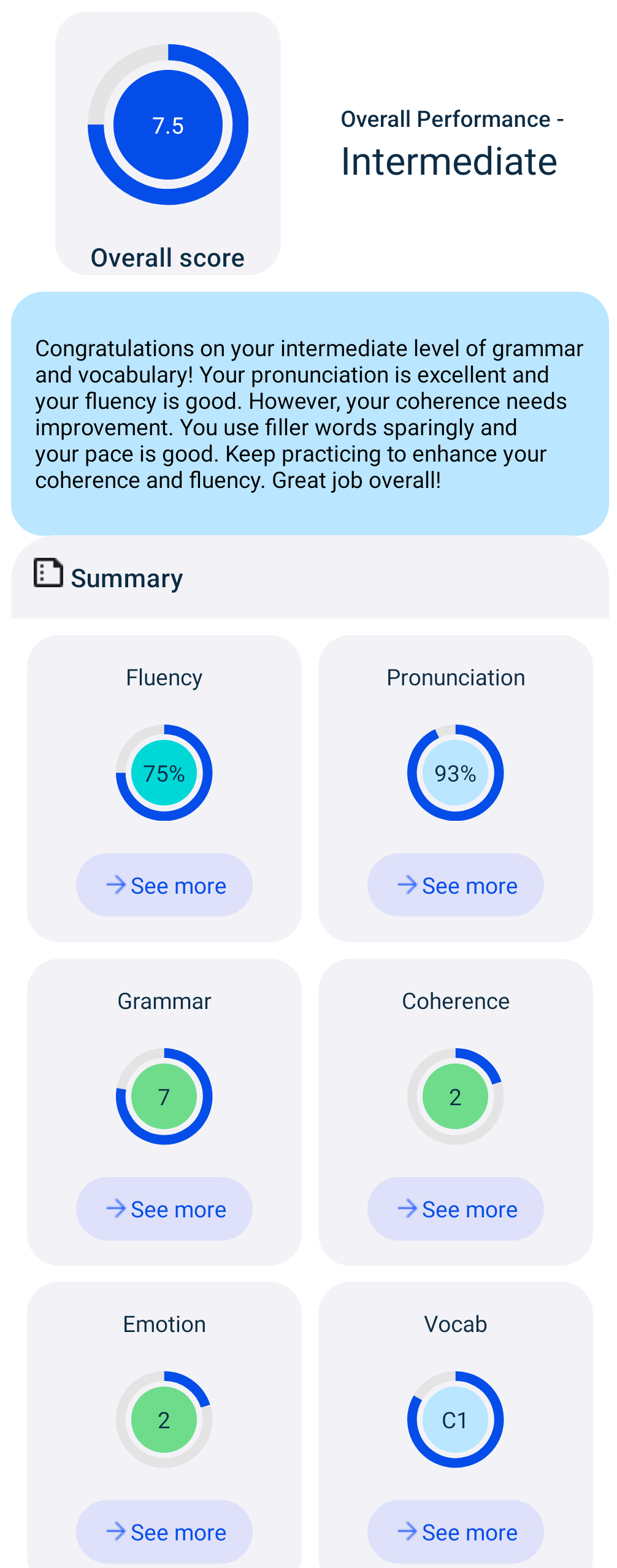}
  \caption{Report Screen}
  \label{fig:report_screen}
\end{minipage}%
\hspace{10pt}
\begin{minipage}{.21\textwidth}
  \includegraphics[width=1\linewidth]{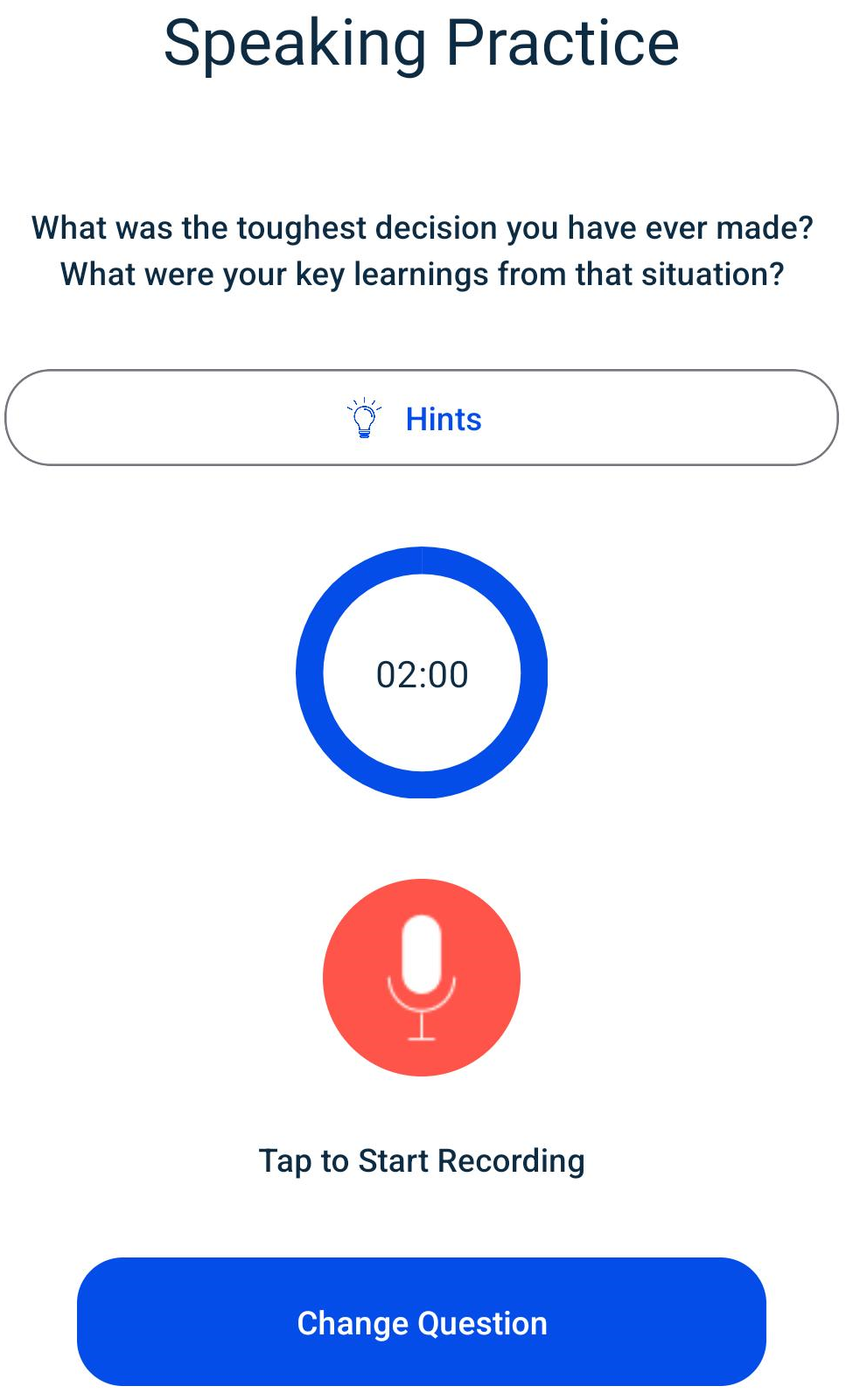}
  \caption{Practice Screen}
  \label{fig:practice_screen}
\end{minipage}
\hspace{10pt}
\begin{minipage}{.21\textwidth}
  \includegraphics[width=1\linewidth]{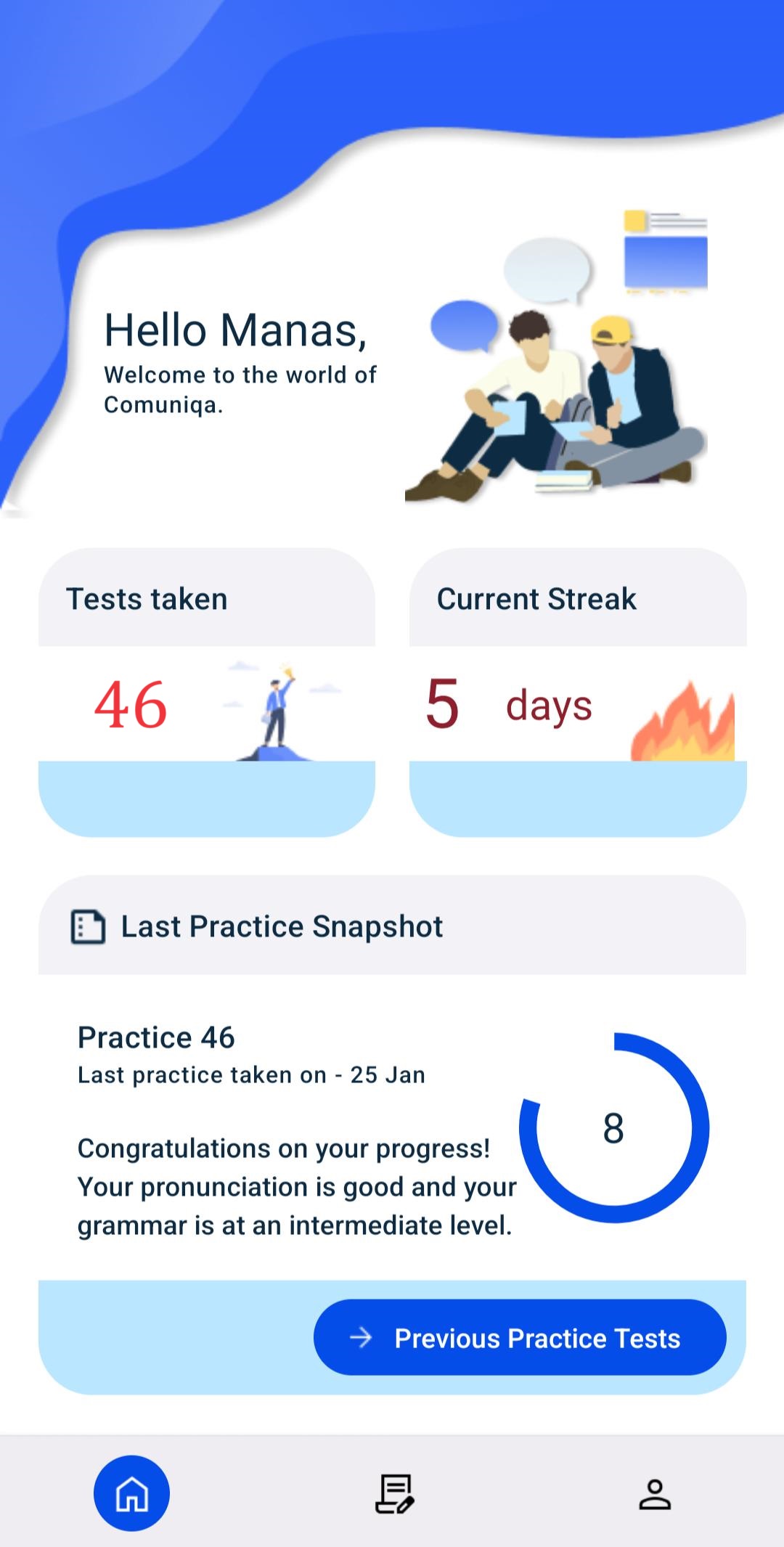}
  \caption{Home Screen}
  \label{fig:home_screen}
\end{minipage}
\hspace{10pt}
\begin{minipage}{.21\textwidth}
  \includegraphics[width=1\linewidth]{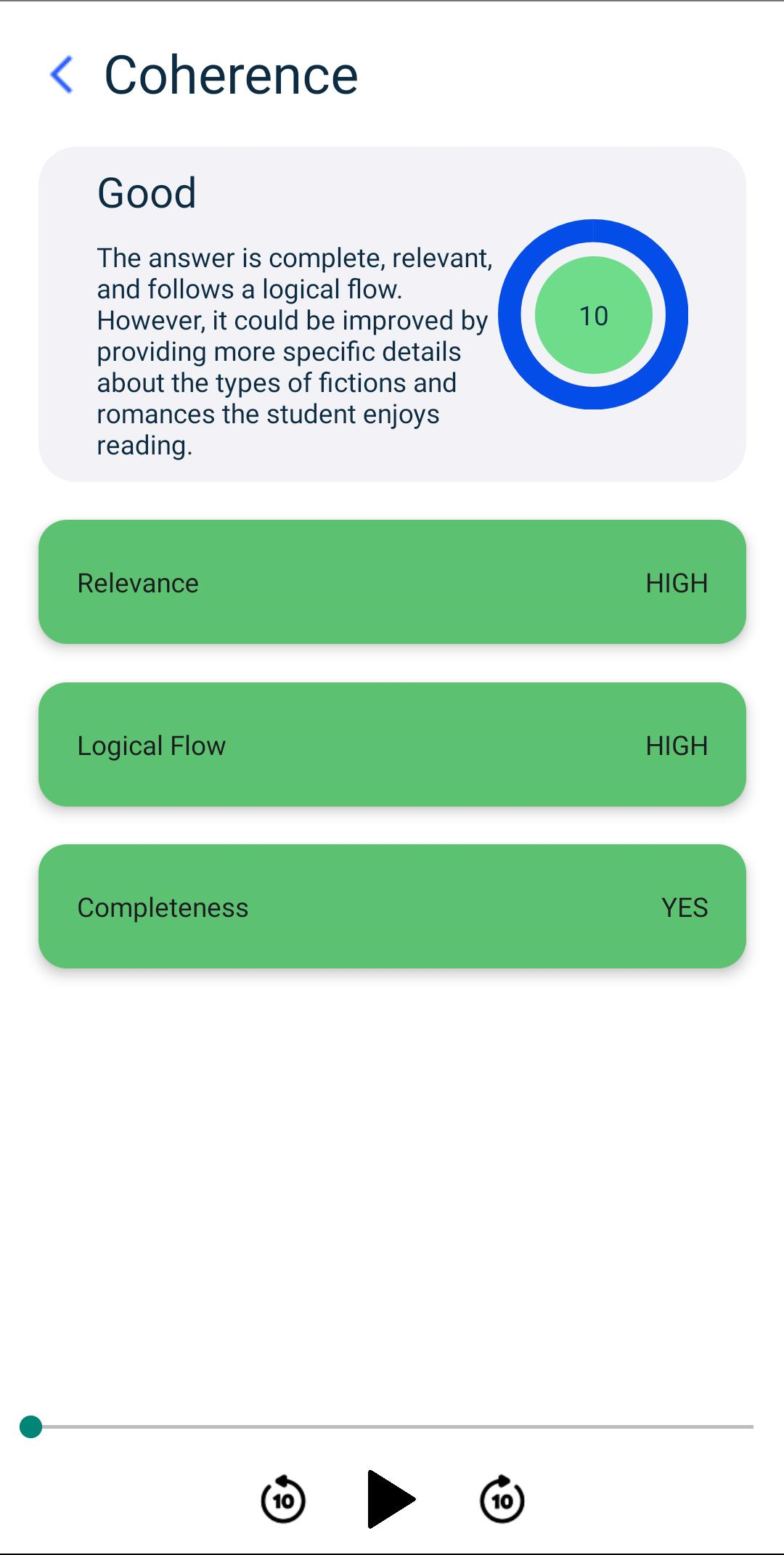}
  \caption{Coherence}
  \label{fig:coherence_screen}
\end{minipage}%

\hspace{10pt}
\begin{minipage}{.19\textwidth}
  \includegraphics[width=1\linewidth]{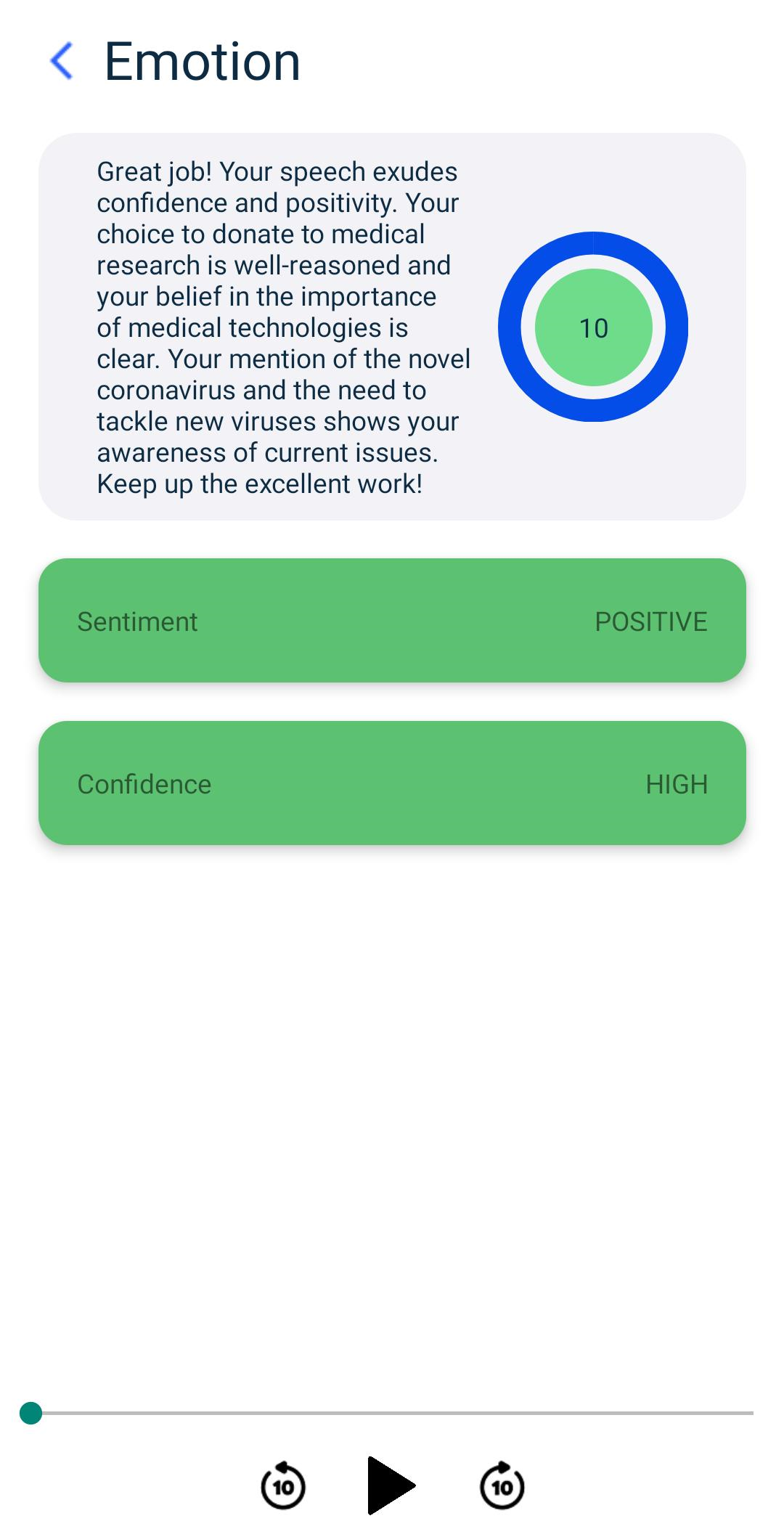}
  \caption{Emotion}
  \label{fig:emotion_screen}
\end{minipage}%
\begin{minipage}{.19\textwidth}
  \includegraphics[width=1\linewidth]{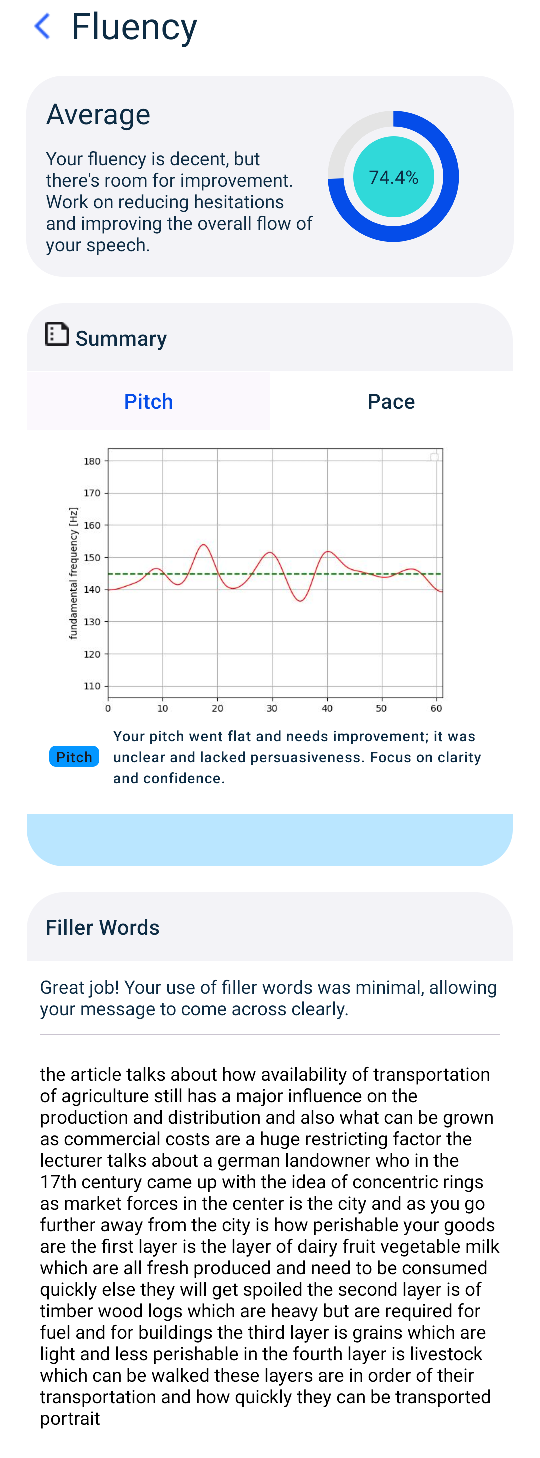}
  \caption{Fluency}
  \label{fig:fluency_screen}
\end{minipage}
\
\begin{minipage}{.19\textwidth}
  \includegraphics[width=1\linewidth]{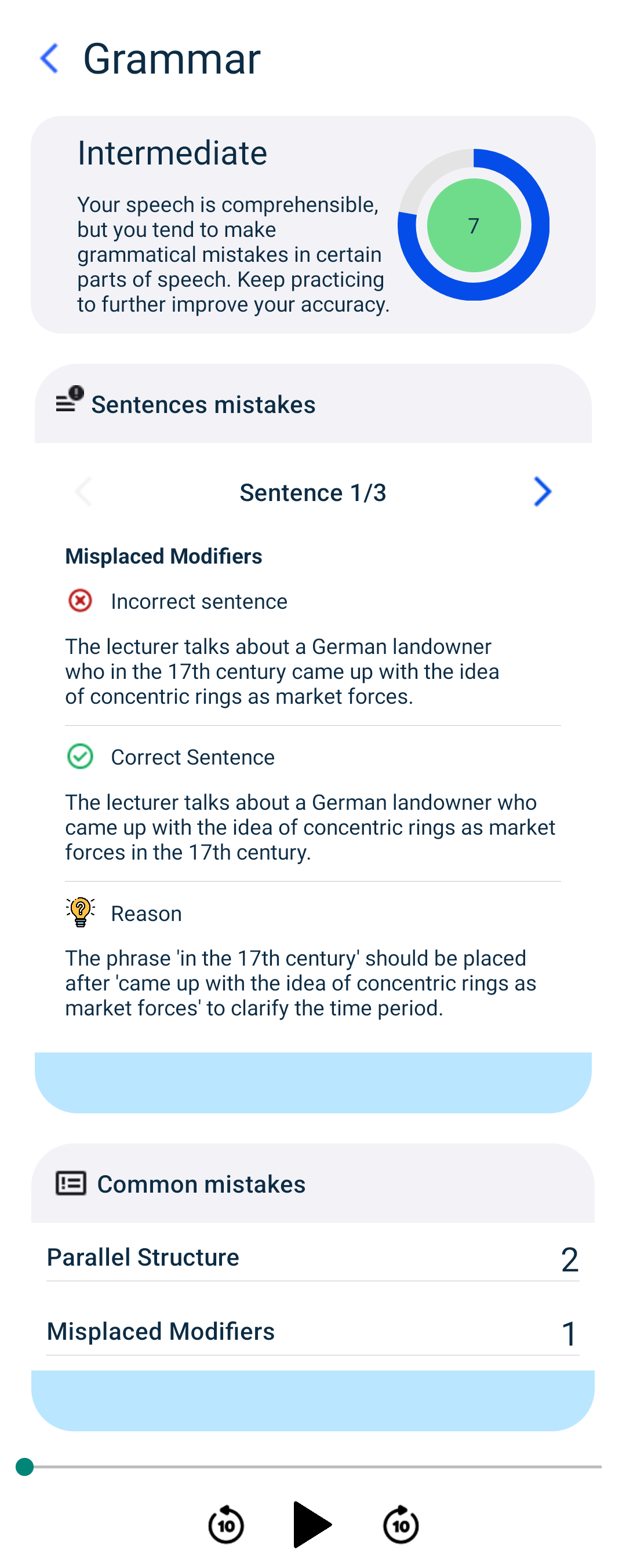}
  \caption{Grammar}
  \label{fig:grammar_screen}
\end{minipage}
\begin{minipage}{.19\textwidth}
  \includegraphics[width=1\linewidth]{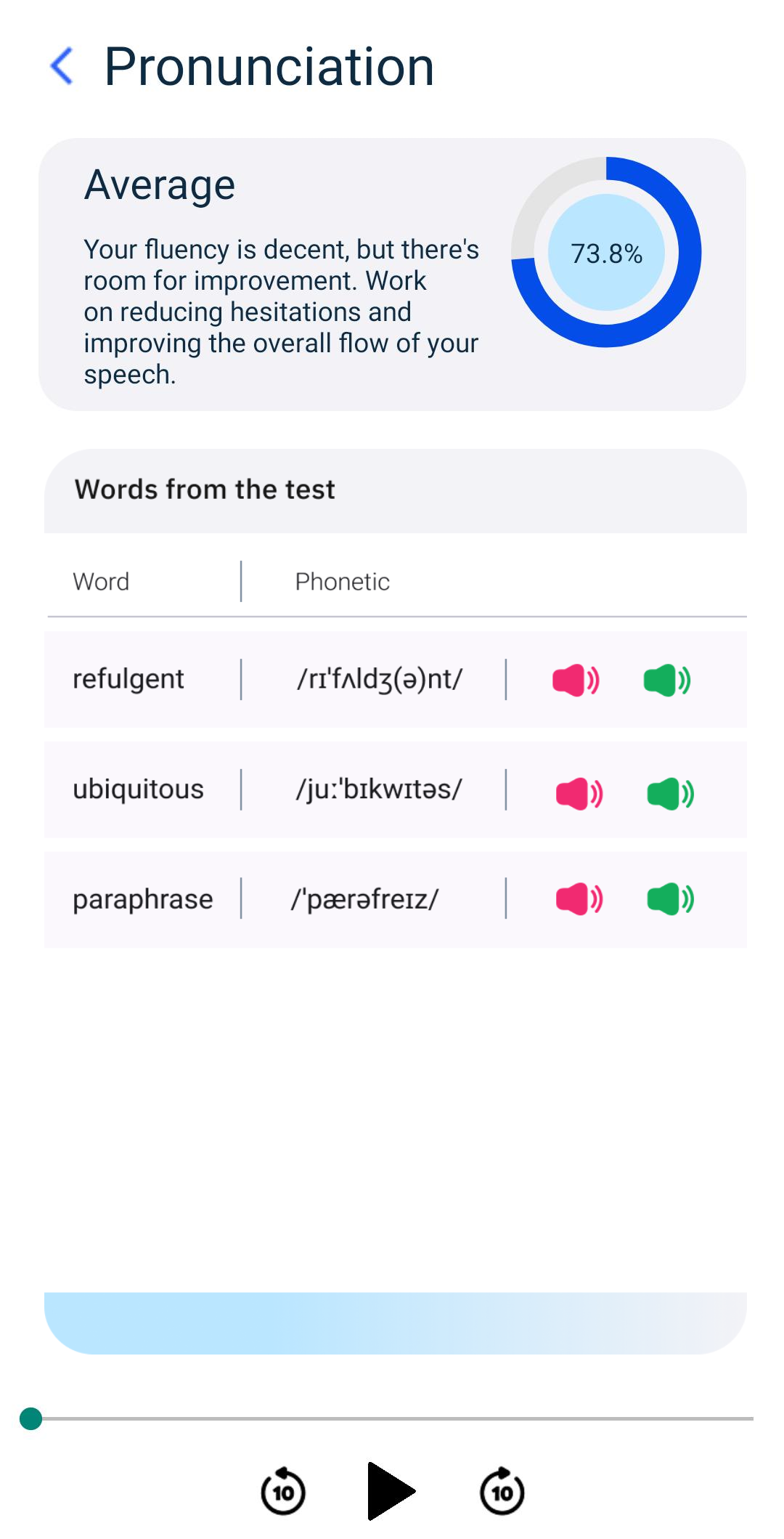}
  \caption{Pronunciation}
  \label{fig:pronunciation}
\end{minipage}
\begin{minipage}{.19\textwidth}
  \includegraphics[width=1\linewidth]{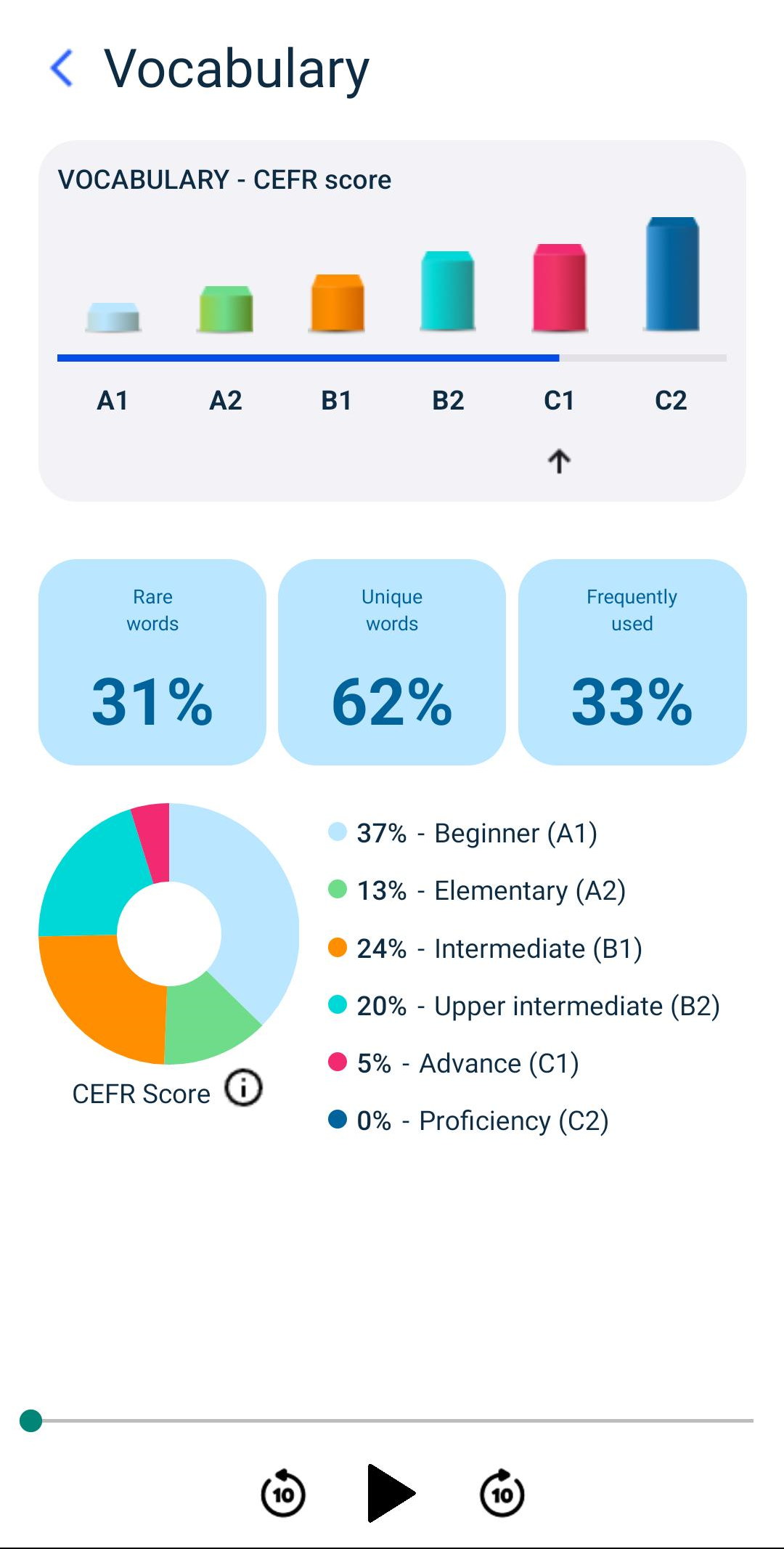}
  \caption{Vocab}
  \label{fig:vocab}
\end{minipage}
\end{figure*}

Once logged in, users can opt to practice speaking for three specific scenarios: 1) IELTS/TOEFL exams, 2) Interviews, and 3) Custom Practice. For the first two scenarios, a pre-filled library of relevant questions is available, while for custom practice, users can input any question of their choice they wish to practice. After selecting a scenario and question, users have 1-3 minutes to record their response. They can change questions and record their answers until satisfied. Upon submission, they receive a detailed analysis of their response based on several parameters listed in Table \ref{tab:app_parameters} and above. A high-level overview report is shown in Figure \ref{fig:report_screen} and users can also delve into a detailed analysis of individual parameters (Figures \ref{fig:practice_screen},  \ref{fig:home_screen}, \ref{fig:coherence_screen}, \ref{fig:emotion_screen}, \ref{fig:fluency_screen} \ref{fig:grammar_screen}, \ref{fig:pronunciation} and \ref{fig:vocab}). The objective was to align the app parameters closely with the attributes evaluated by human experts. Additionally, users can review their past practice sessions in the history section and view the transcript and playback recorded audio for each session.

\begin{table}[ht]
\centering
\begin{minipage}{\columnwidth}
    \centering
    \footnotesize
    \begin{tabular}{|c|p{3.5cm}|}
        \hline
        \textbf{Parameters}  & Detailed Attributes\\
        \hline
        \textbf{Overall Summary} & Numerical Score, Overall Summary and Ideal Response\\ \hline 
        \textbf{Vocabulary} & CEFR Score, Distribution of Word across CEFR levels\\ \hline 
        \textbf{Fluency} & Pace, Pitch, Filler Words and Awkward Pauses\\ \hline 
        \textbf{Emotion} & Sentiment and Confidence\\ \hline 
        \textbf{Pronunciation} & Mispronounced Words and their correct pronunciations\\ \hline 
        \textbf{Coherence} & Relevance, Logical Flow and Completeness\\ \hline 
        \textbf{Grammar} & Grammatical errors and corresponding correct sentence structures\\
        \hline 
    \end{tabular}
    \caption{Parameters used in the feedback provided by LLM-based application}
    \label{tab:app_parameters}
\end{minipage}
\vspace{1em}
\end{table}

\section{Methodology}\label{sec:methods}















\subsection {Recruitment}

\subsubsection{Participant Recruitment}
The recruitment process for our study aimed to assemble a representative and diverse cohort from a targeted population that comprised students with an interest in improving their speaking skills. 
To ensure a diverse participant pool, we shared a Google form with all the students via the university emailing lists. The survey (Appendix \ref{sec:appendix_questions}) captured essential information, encompassing demographic details, academic standing, native or primary language, the main reason for seeking improvement in speaking skills, and a self-evaluation of current English-speaking proficiency on a Likert scale from 1 (Beginner) to 5 (Expert). The survey format included five multiple-choice questions and one open-ended question. 

Out of the 160 responses we received, we selected 36 participants with an average age of 19.4 ($\sigma$ = 1.9). While selecting, we made sure to incorporate gender diversity (28 male and 8 female participants), linguistic diversity (8 non-English native languages) and diversity in linguistic proficiency (self-reported). Demographic information about each participant is mentioned in Appendix \ref{sec:appendix_demographics} (Table \ref{table:participant_demographics}).
We also selected 4 additional participants for a pilot study prior to the main study.
Each participant was remunerated within a range of Rs. 200 to Rs. 400. 
This remuneration aimed to recognize the time and effort invested by participants, aligning with ethical considerations and fostering a sense of appreciation for their invaluable contributions to the research study.

\subsubsection{Expert Recruitment}
For the study, we recruited three (N=3) experts to conduct all the expert-led sessions with the participants. The task of selecting these experts was entrusted to a third-party organization, EnglishYaari \cite{EnglishYaari}, popular online platform having several certified English tutors providing personalized 1-1 sessions. The chosen tutors brought varied educational and professional experiences, including roles as educators, graduate/postgraduate students, and working professionals. Each expert held language certifications like CELTA, TEFL, or TESOL, ensuring a high standard of proficiency and teaching quality. The feedback from experts, which carefully analysed the participants' performances, was thoroughly recorded and organized, consisting of attributes similar to those provided by the LLM-based application (Table \ref{tab:app_parameters}). 

\subsection {Study Design}\label{sec:study_design}
The selected participants underwent three phases in the study: 1) pre-study interview before the study commenced, 2) participated in a two-week study, and 3) post-study interviews after the study's completion. The 36 participants were further divided into three equally-sized groups of 12 individuals using random sampling. All participants sat in a proctored environment with reliable Wifi connection and silent surroundings. During the study, 2 participants dropped out.
\begin{itemize}[leftmargin=*]

    \item \textbf{Group 1 (Comuniqa app Only):} Participants in this group participated in learning sessions conducted through the Comuniqa app, with each session lasting approximately 20 minutes. During each session, participants completed two practice exercises on the Comuniqa app and dedicated time to reviewing the results and feedback provided by app. In total, the participants completed two sessions amounting to 4 speaking practice tests (N=4) over two weeks.

    \item \textbf{Group 2 (Expert Only):} Participants in this group attended expert-led sessions on the EnglishYaari platform, each lasting around 30 minutes. The experts crafted various speaking practice exercises and provided real-time feedback to the participants during the sessions. 
    Each participant completed a total of four sessions with the experts (N=4).

    \item \textbf{Group 3 (Comuniqa app and Expert):} Participants in this group experienced a hybrid learning approach alternating between sessions with human experts and the Comuniqa App based on their preferences.
    Specifically, participants attended two sessions with experts and engaged in two sessions using the Comuniqa app. The expert sessions, conducted online for 30 minutes with certified tutors from the EnglishYaari platform, were complemented by one practice test per session on the app. Each session on the application took approximately 15 minutes, including the time to record and analyze the feedback and reviews provided by the app. Overall, in this group each participant completed a total of two practice tests (N=2) on the app and participated in two expert-led sessions (N=2). 
\end{itemize}

We divide the participants into the above mentioned three groups for understanding the impact of tutoring by LLM-based system (Comuniqa) and human experts on learners' speaking skills. This approach will allow us to dissect and isolate the individual impact of Comuniqa, human experts, and their combined impact on participants' speaking proficiency. Through separate groups, we aim to gather insights into user preferences, discerning participants' inclinations towards various learning approaches. 
Additionally, this design facilitates the identification of users who may potentially want to switch between the LLM-based tutoring and human tutoring, offering valuable insights into dynamic user preferences over the study duration.
It also helps us mitigate potential confounding factors, contributing to the study's internal validity by minimizing the risk of contamination between different tutoring methods.

Pilot study was used to assess the course and direction of the research. Our pilot study simulated the main study (described above) among the additionally selected four participants. We incorporated the feedback and recommendations from the users in the pilot study to streamline our main study.

\section{Evaluations}\label{sec:evaluation}





\subsection{Usage and Engagement}

We had a total of 34 participants who gave a total of 64 tests on the Comuniqa app and 70 user sessions with human experts. The average engagement time per active user on the mobile app was 26m 26s with an average engagement time of 11m 54s per session. The average time for each session with experts was around 29 minutes. Screens for grammar and vocabulary reports had the highest average engagement time of 2 minutes 5 seconds and 1 minute 46 seconds per user, respectively. On the coherence and emotion screens, users spent the least time with 15 seconds and 17 seconds on average for each screen, respectively.

\subsection{Quantitative Results}
\begin{figure*}
\centering
\label{}
\begin{minipage}{.5\textwidth}
  \includegraphics[width=1\linewidth]{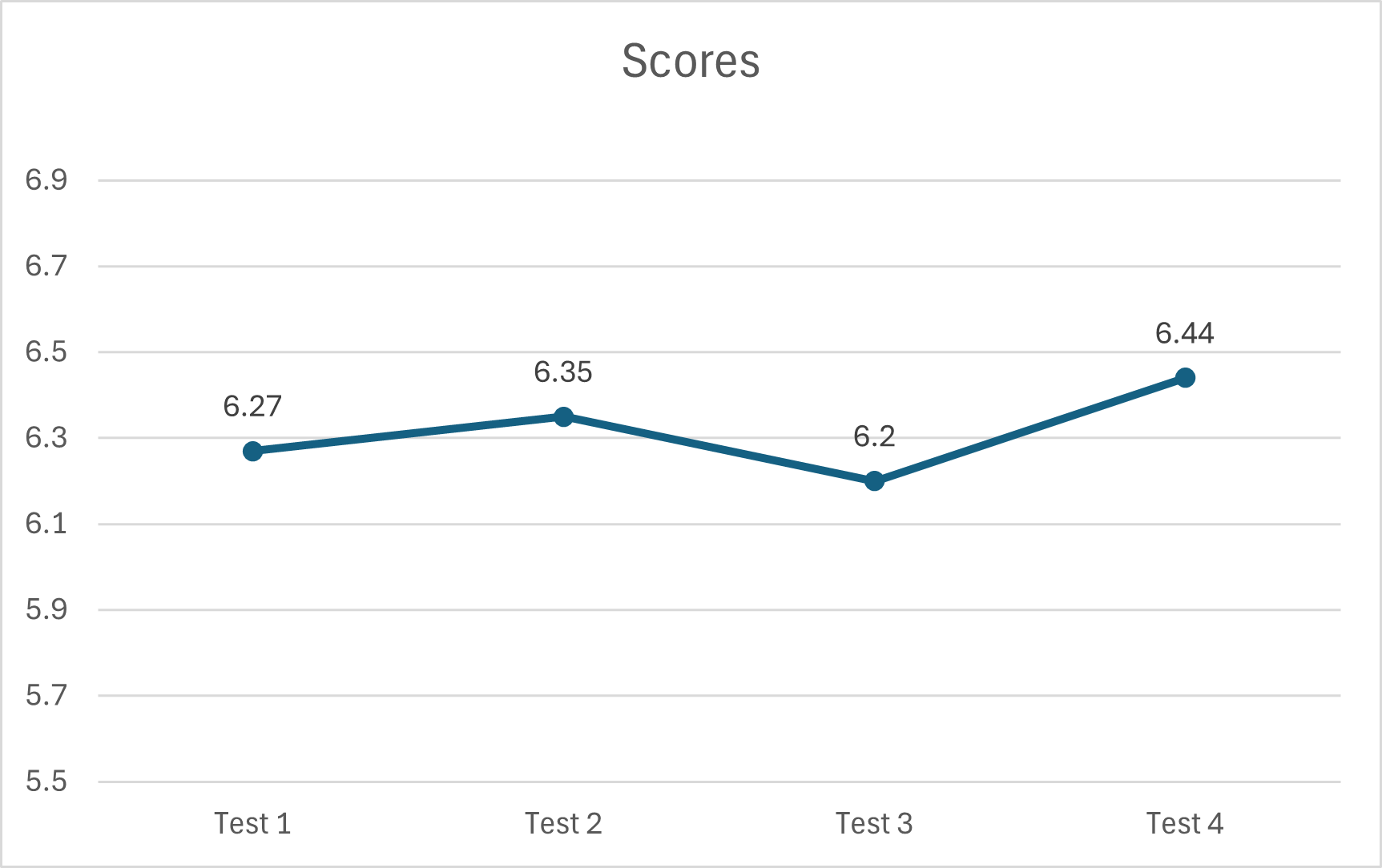}
  \caption{Average session scores for Group 1 users}
  \label{fig:avg_scores_plot_group1}
\end{minipage}%
\begin{minipage}{.5\textwidth}
  \includegraphics[width=1\linewidth]{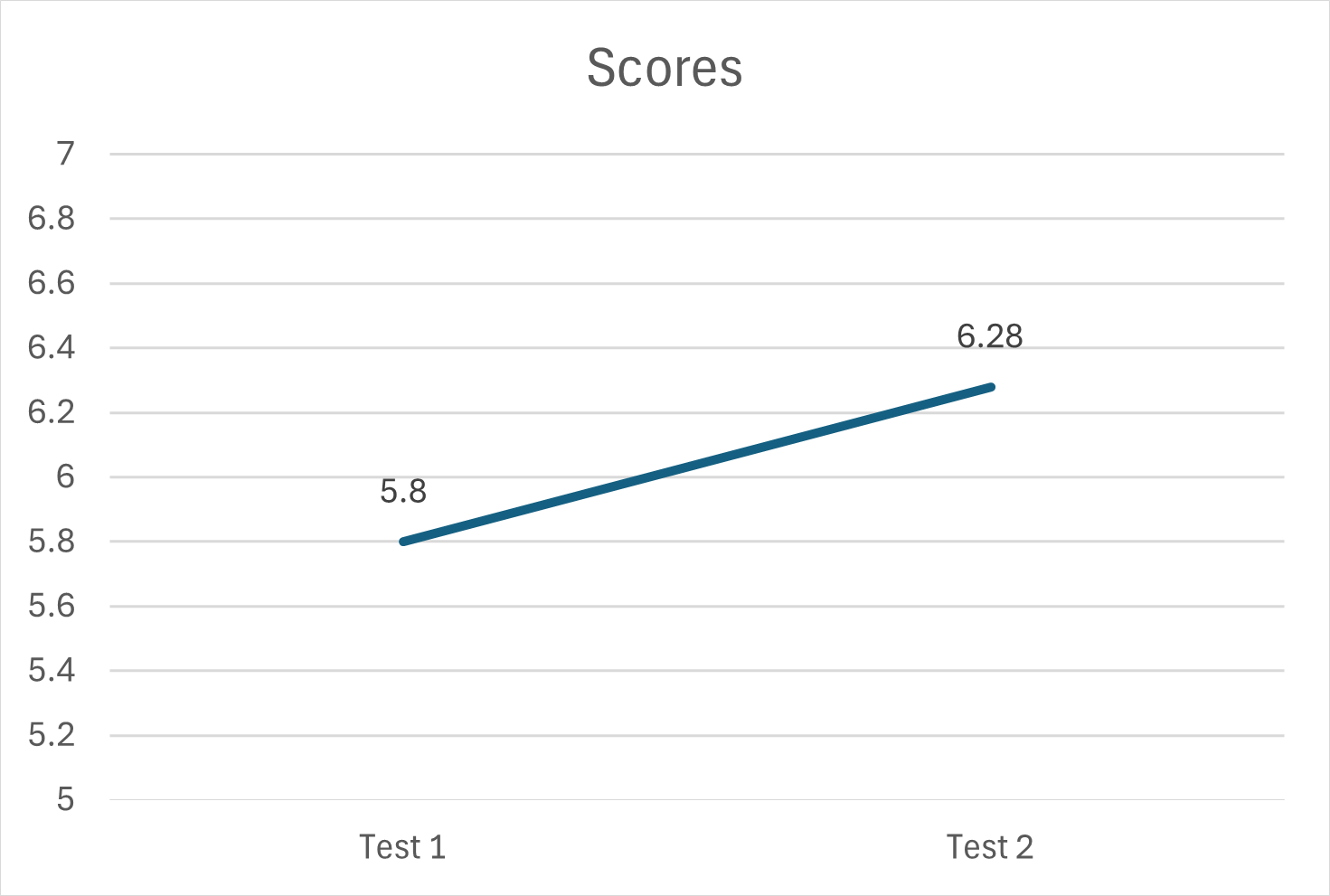}
  \caption{Average session scores for all Group 3 users}
  \label{fig:avg_scores_plot_group3}
\end{minipage}
\begin{minipage}{.5\textwidth}
  \includegraphics[width=1\linewidth]{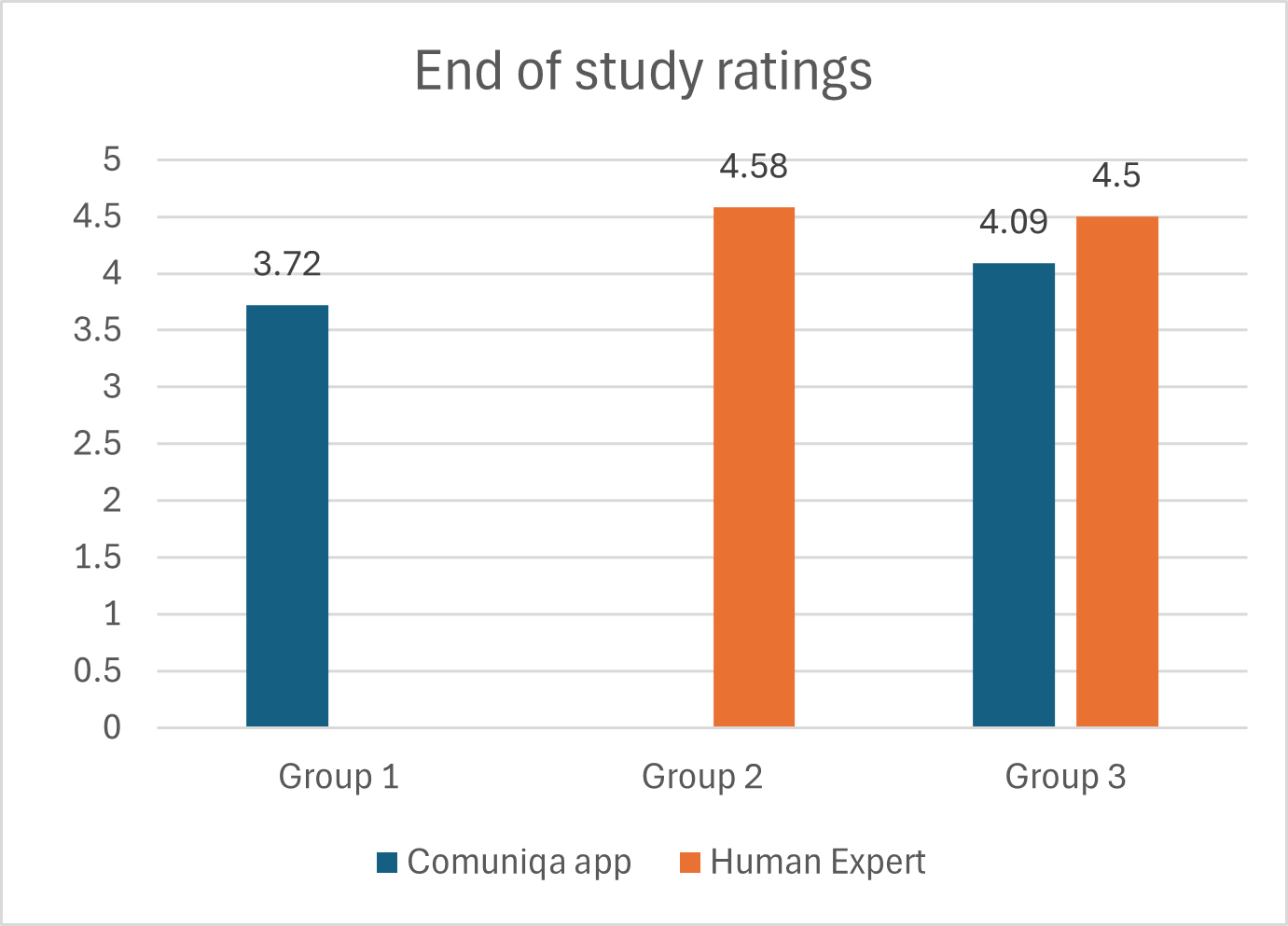}
  \caption{Average ratings for all three groups}
  \label{fig:ratings}
\end{minipage}%
\begin{minipage}{.5\textwidth}
  \includegraphics[width=1\linewidth]{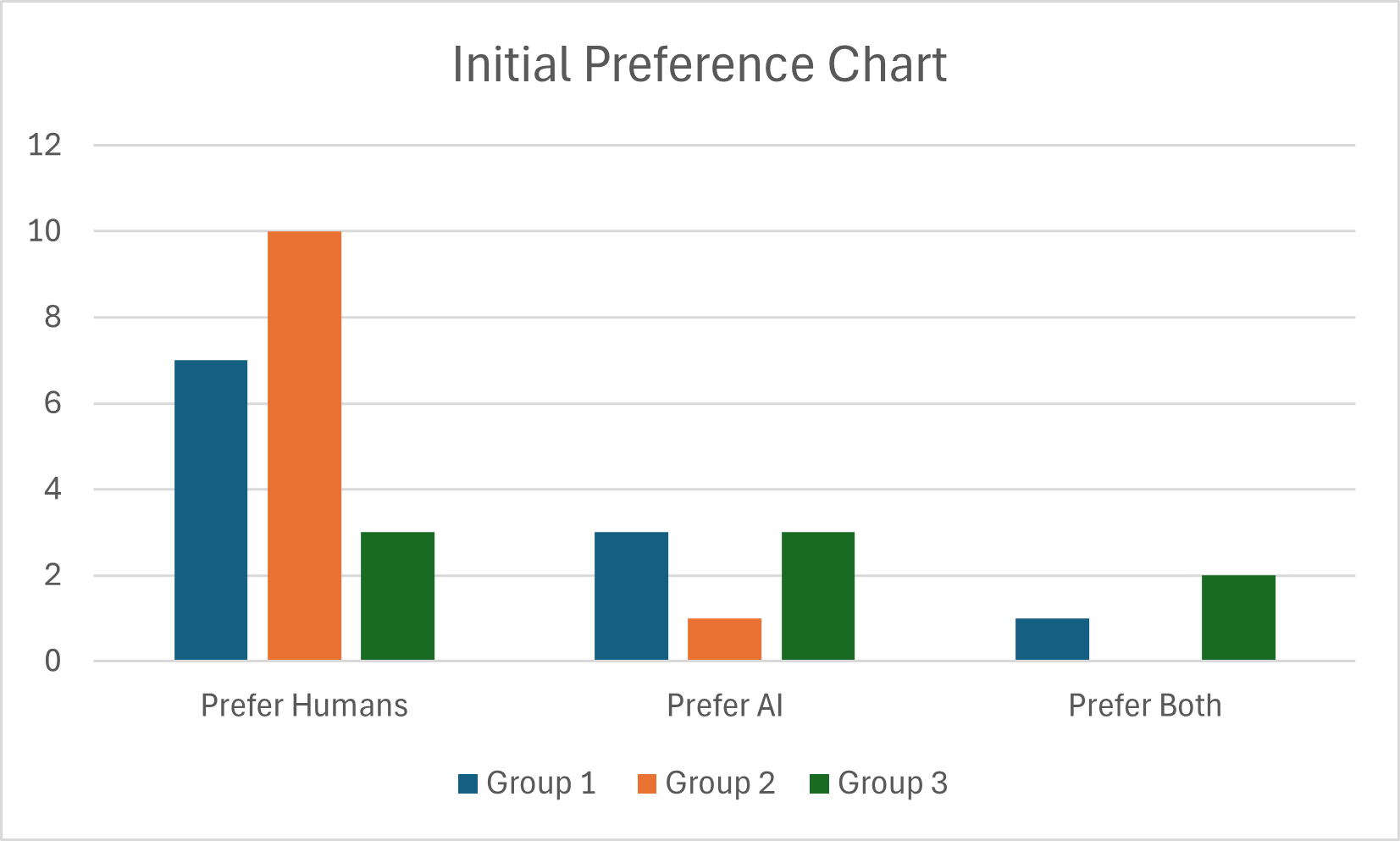}
  \caption{Initial Preference between AI tools or Human Experts}
  \label{fig:InitialPreferenceChart}
\end{minipage}
\end{figure*}

We did not observe any noticeable difference in the average or individual student scores across multiple app sessions conducted using the Comuniqa App (Figure \ref{fig:avg_scores_plot_group1} and Figure \label{fig:avg_scores_plot_group3}). This can be attributed to the fact that all the sessions were conducted within a span of two weeks, during which it was not anticipated that the participants would demonstrate substantial enhancement in their speaking skills. 

Each participant was asked to rate their overall experience with the Comuniqa app and/or the human expert on a Likert scale of 1 (Poor) to 5 (Excellent). The average rating provided by participants for their overall experience with the Comuniqa app in Group 1 and Group 3 is 3.72 $\pm$ 0.88 and 4.58 $\pm$ 1.08 respectively. The average rating for the human experts in Group 2 and Group 3 is 4.09 $\pm$ 0.60 and 4.5 $\pm$ 0.95 respectively. Overall, we see that the ratings are slightly higher for human experts compared to the Comuniqa app (Figure \ref{fig:ratings}). 

\begin{figure*}
    \center
    \includegraphics[width=0.4\textwidth]{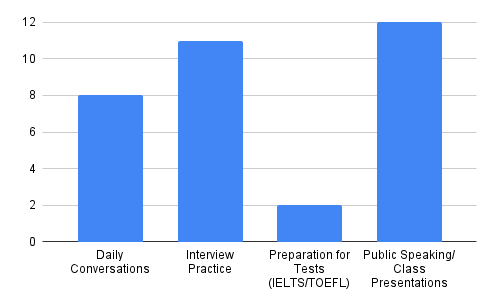}
    \caption{Motivation for improving speaking skills}
    \label{fig:motivation}
\end{figure*}

In terms of participants' motivation for improving their speaking skills, the users wanted to improve their speaking skills for various reasons such as improving day-to-day conversations, interview preparation, standard language tests (IELTS/TOEFL), public speaking and class presentations (Figure \ref{fig:motivation}). We plan to conduct a more extended study to validate our findings further.

\subsection{Qualitative Findings}

We conducted semi-structured interviews as part of the pre-study and post-study, which were then transcribed and coded using thematic analysis techniques, as outlined by Braun and Clarke \cite{thematicAnalysis}. In the pre-study interviews, we gathered the motivations of users for wanting to improve their speaking skills, any prior experiences with human experts or apps to improve spoken English and their initial preference of human experts vs AI for practising spoken English. Through the post-study interview, we extracted the participant's views on the session and on the helpfulness of their session with the app and/or with the human expert. We also wanted to understand the key benefits and limitations of each of these and if the user's opinion changed from their initial perception of AI vs human experts (Figure \ref{fig:InitialPreferenceChart}). 

\begin{table}[!ht]
    \centering
    \small
    \begin{tabular}{|p{2.2cm}|p{6cm}|}
    \hline
        \textbf{AI} - Accuracy \& Trust & \textit{"I'm not sure about the accuracy of the test scores because I don't know what were the metrics used."} [P9, G1] \\ \hline
        \textbf{AI} - Accessibility \& Convenience & \textit{The app gave reports quickly within a couple of minutes and accessible as I can practice any time of the day.} [P2, G1] \\ \hline
        \textbf{AI} - Non-Judgmental & \textit{"No inferiority complex or shyness when talking to the app, so I could talk freely without any fear of judgement"} [P3, G1] \\ \hline
        \textbf{Human Experts} - Interaction and Motivation & \textit{"The interaction with the expert was very relaxed, felt natural and more easy, therefore I performed better"} [P31, G3] \\ \hline
        \textbf{Human Experts} - Emotions \& Empathy & \textit{"Humans are much better at understanding than AI; they have real emotions, AI uses algorithms and don’t have common sense or
feelings while humans will usually know what the person is trying to say and understand where he/she is lacking."} [P29, G3] \\ \hline
        \textbf{Human Experts} - In-depth critique & \textit{"The expert knew my weaknesses exactly, when I hesitate and I am lacking proper vocabulary to express myself."} [P17, G2] \\ \hline
    \end{tabular}
    \caption{Main themes emerging from our qualitative analysis. First three themes correspond to AI and the next three themes correspond to human experts.}
    \label{tab:quotes}
    \vspace{-2em}
\end{table}

\subsubsection{\textbf{Perceived Benefits of Human Experts}}
When asked about their initial preferences between AI and tools and human experts, a majority (N=22) of the participants strongly preferred humans for practicing English because humans have emotions and the ability to understand and empathize that AI tools cannot yet recreate. P26 stated that \textit{“Humans are much better are understanding than AI, they have real emotions}”. P20 echoed the same sentiment, \textit{“AI use algorithms and don't have common sense or feelings while humans will usually know what the person is trying to say and understand where he/she is lacking”}.  

Several participants also stated that seeing other humans speak fluently motivates them, influencing their own speaking abilities. \textit{“AI doesn’t sound pleasing. When I see other humans speaking well, it's good to listen to them, I admire them, and I feel motivated to speak like them.”} - [P5] 

P33 also pointed out that since there was real human in front of the screen,\textit{ “the one to one session was very natural, the expert could understand me better and similarly I could understand him better.”}

\subsubsection{\textbf{Perceived Limitations of Human Experts}}
Participants interacting with human experts admitted that scheduling a call based on availability was a real drawback. "Fixing a common time with the human expert was hard. Sometimes they used to be available when I wasn’t and often the other way round." - [P15] \newline
Certain participants who were not confident in their speaking skills were hesitant to take the sessions with human experts fearing judgement.

\subsubsection{\textbf{Perceived Benefits of AI}}
Out of the total (N=7) participants that preferred AI tools, a majority (N=5) of them reasoned that there is no judgement or bias while talking to AI, and you can freely record your speech without feeling shy or having any inferiority complex which can occur in front of human experts. 

From users who interacted with the app, several users (N=5) found the app and the generated report \textit{“accurate and quite helpful for me to inspect where I lack”} -[P4].  Two users mentioned that they found it very beneficial to read through their actual transcript to self-analyse the content and vocabulary of their speech.  \textit{“The app transcribed my entire speech, so I can go back and actually see what I actually said”} -[P11].

Some users liked that the app analyzed several different parameters, and gave them structured feedback and a score, especially useful for exams like TOEFL and IELTS. 
Users also praised the accessibility and wide range of practice questions on the app. Participants also added that they could \textit{“talk to the app freely, without any fear of judgement”}, as we observed from the pre-study interviews as well. P10 shared how such an app would be highly beneficial especially for students from underprivileged backgrounds. \textit{“In my coaching there were a lot of kids from underprivileged backgrounds. The teacher used to take in students from nearby villages and give them a platform to study. These kids didn't have proper English coaching, so they used to sit on computers and have those programs where they would listen and then speak. But these systems were very basic and there wasn’t AI implemented in those systems. Your app would be good in that sort of English tutor or English tutor websites.”}




\subsubsection{\textbf{Perceived Limitations of AI}}
As mentioned earlier, users really felt a lack of human element in the app, \textit{"it was very mechanical, I feel I need a human connect to learn and understand where I lack".} 
For participants interacting with the app, atleast 3 users added that they didn't not trust the judgement of the app especially the results generated by AI evaluation. P29 felt that, \textit{“sometimes I felt the score given by the app was wrong. I feel I performed better in the last test, but I got a low score in fluency, which should have been more.”} 
P20 stated, \textit{“AI has no emotion and support and you can't have a general chit chat. I was unwell for few days between sessions so the mentor asked me how I was doing. In each session the beginning we discussed random stuff for a few minutes, something that I will never get with AI.”} 
Others mentioned that the app did not have follow up questions, and faced issues interacting with the app interface and needed more transparency on how the scores were calculated.
A participant also pointed out that they'd prefer if metrics such as pace and pitch would appear as live feedback during the session.
\subsubsection{Changes in User Preferences}
Participants preferences post the interaction with the app or experts did not see any major changes. A singular participant from Group 2 who stated that his initial preference was AI tools, later went on to change his opinion post interaction with the human expert, stating that \textit{“human experts are much better than the online tools because they can identify my actual weaknesses. They don't need any explanation or context; they can just observe from your speaking skills where exactly you are lagging and since they have accumulated a lot of experience over time that experience is really helpful”}. 
Similarly, in group 3, a participant admitted that the human expert was much better, \textit{"the human expert gave firsthand feedback that was easy to implement in the following sessions, it was more actionable than the pointers the app gave"}.
P4, who had initially preferred human expert later switched to preferring the AI app, \textit{"the feedback from the humans was very generic, statistics like pace, pitch and number of grammatical errors that the app provides are much more helpful"}.
\section{Discussion}\label{sec:discuss}
This study aimed to explore the efficacy of LLM-based systems in supporting language learning, specifically focusing on speaking skills. This work is set against the backdrop of the rising popularity of modern LLM-based language learning applications, evidenced by hundreds of such apps on Google Playstore and Apple Appstore. We developed Comuniqa, an LLM-based application that evaluates user input and provides feedback on various standard parameters. Additionally, a subset of users received feedback from human experts alongside the Comuniqa app, allowing us to understand preferences and the nuances of each modality. 

\textbf{Answering RQ1,} our preliminary analysis suggests a strong preference among participants for human experts, with empathy being a significant factor, and interactions being perceived as more natural. While users acknowledged the AI system's accuracy and comprehensiveness, most continued to favor learning from human experts, with only a few (N=3) changing their preference post-study interviews, 2 from AI to human and 1 from human to AI based system. A recurrent theme was that current AI systems cannot match human cognition, both in terms of accuracy and in providing emotional support and empathetic responses. Interestingly, participants expected the AI system to exhibit empathy, illustrating the anthropomorphization of AI systems — a well-studied phenomenon \cite{machineAndMindlessness}. This contrasts with the perceived benefit of AI systems being non-judgmental and unbiased. Hence, while the absence of judgment was seen as a benefit, the lack of emotional response was perceived as a potential flaw. This presents an intriguing design dilemma due to the contrasting expectations.

\textbf{Answering RQ2}, human experts are pivotal in any AI-based system we design, not only for ensuring accuracy and correctness but also for providing the 'human touch', which appears to be a necessity. While we highlight the complexities and nuances of integrating AI in language learning and the indispensable role of human interaction and understand that personalized one-on-one coaching with experts is most optimal, it is often prohibitive due to constraints like time, availability of experts, fear of bias, and cost. We believe thoughtfully designed systems will be able to encompass the benefits of both and this will be the scope of future work.
\vspace{-1em}

\section{Limitations and Future work}\label{sec:discuss}

The study confronts a few limitations that merit consideration in the interpretation of its findings. Firstly, the duration of the study which spanned over a relatively short two weeks, imposes constraints on the comprehensive evaluation of critical aspects such as engagement, learning outcomes, and retention rates. The brevity of the study restricts the attainment of statistical significance in assessing these critical aspects of AI-assisted learning. Additionally, the participant pool predominantly comprised engineering students within a tier-1 city and a specific age group. To enhance the generalizability of findings, future work could involve diversifying the participant population further. Secondly, the absence of a robust framework addressing non-verbal communication skills within Comuniqa app represents a notable limitation. The app underscores the need for a more comprehensive coverage of non-verbal cues, recognizing their significance in fostering the holistic development of communication abilities among learners. This limitation prompts avenues for future research to explore and integrate evaluation of non-verbal communication.
Furthermore, P34 felt that the app lacks a conversational component. This limitation restricts the exploration of the full spectrum of interactive language learning experiences involving dynamic conversational scenarios. Future work could involve the development of conversational tool into Comuniqa app to broaden the scope of interactive language learning.

An avenue for future research involves the incorporation of experts collaborating synergistically with AI systems. Establishing a continuous feedback cycle between human experts and AI platforms could enhance the learning experience and contribute to refining educational interventions. These identified limitations emphasize the ongoing need for investigations and improvements in seamlessly integrating AI technologies into educational contexts.
\section{Conclusion}\label{sec:conclusion}
In this paper, we presented a novel LLM-based platform for improving speaking skills. We evaluated the strengths and weaknesses of this system by conducting a preliminary user study involving 34 participants. We divided the participants into three groups: those who only use our LLM-based platform for improving speaking skills, those who only interact with human experts and those who interact with both the LLM-based platform as well as the human experts. Our early findings suggest that while LLMs are good at providing strong quantitative feedback, they cannot match the cognitive abilities of human experts, both in terms of accuracy and empathy. Nevertheless, our proposed LLM-based platform represents a step towards achieving \emph{Sustainable Development Goal 4: Quality Education} by providing a valuable learning tool for individuals who may not have access to human experts for improving their speaking skills.

\bibliographystyle{ACM-Reference-Format}
\bibliography{references}


\begin{thebibliography}{37}


\ifx \showCODEN    \undefined \def \showCODEN     #1{\unskip}     \fi
\ifx \showDOI      \undefined \def \showDOI       #1{#1}\fi
\ifx \showISBNx    \undefined \def \showISBNx     #1{\unskip}     \fi
\ifx \showISBNxiii \undefined \def \showISBNxiii  #1{\unskip}     \fi
\ifx \showISSN     \undefined \def \showISSN      #1{\unskip}     \fi
\ifx \showLCCN     \undefined \def \showLCCN      #1{\unskip}     \fi
\ifx \shownote     \undefined \def \shownote      #1{#1}          \fi
\ifx \showarticletitle \undefined \def \showarticletitle #1{#1}   \fi
\ifx \showURL      \undefined \def \showURL       {\relax}        \fi
\providecommand\bibfield[2]{#2}
\providecommand\bibinfo[2]{#2}
\providecommand\natexlab[1]{#1}
\providecommand\showeprint[2][]{arXiv:#2}

\bibitem[spr({[n.\,d.]})]%
        {springerEffectsReciprocal}
 \bibinfo{year}{[n.\,d.]}\natexlab{}.
\newblock \bibinfo{title}{{E}ffects of reciprocal peer feedback on {E}{F}{L} learners’ communication strategy use and oral communication performance - {S}mart {L}earning {E}nvironments --- link.springer.com}.
\newblock \bibinfo{howpublished}{\url{https://link.springer.com/article/10.1186/s40561-018-0061-2}}.
\newblock
\newblock
\shownote{[Accessed 26-01-2024]}.


\bibitem[Ashby et~al\mbox{.}(2023)]%
        {Ashby2023}
\bibfield{author}{\bibinfo{person}{Trevor Ashby}, \bibinfo{person}{Braden~K. Webb}, \bibinfo{person}{Gregory Knapp}, \bibinfo{person}{Jackson Searle}, {and} \bibinfo{person}{Nancy Fulda}.} \bibinfo{year}{2023}\natexlab{}.
\newblock \showarticletitle{{Personalized Quest and Dialogue Generation in Role-Playing Games: A Knowledge Graph- and Language Model-based Approach}}. In \bibinfo{booktitle}{\emph{Conference on Human Factors in Computing Systems - Proceedings}}. \bibinfo{publisher}{Association for Computing Machinery}, \bibinfo{pages}{20}.
\newblock
\showISBNx{9781450394215}
\urldef\tempurl%
\url{https://doi.org/10.1145/3544548.3581441}
\showDOI{\tempurl}


\bibitem[Becker et~al\mbox{.}(2023)]%
        {becker2023programming}
\bibfield{author}{\bibinfo{person}{Brett~A. Becker}, \bibinfo{person}{Paul Denny}, \bibinfo{person}{James Finnie-Ansley}, \bibinfo{person}{Andrew Luxton-Reilly}, \bibinfo{person}{James Prather}, {and} \bibinfo{person}{Eddie~Antonio Santos}.} \bibinfo{year}{2023}\natexlab{}.
\newblock \showarticletitle{Programming Is Hard - Or at Least It Used to Be: Educational Opportunities and Challenges of AI Code Generation}. In \bibinfo{booktitle}{\emph{Proceedings of the 54th ACM Technical Symposium on Computer Science Education V. 1}} (<conf-loc>, <city>Toronto ON</city>, <country>Canada</country>, </conf-loc>) \emph{(\bibinfo{series}{SIGCSE 2023})}. \bibinfo{publisher}{Association for Computing Machinery}, \bibinfo{address}{New York, NY, USA}, \bibinfo{pages}{500–506}.
\newblock
\showISBNx{9781450394314}
\urldef\tempurl%
\url{https://doi.org/10.1145/3545945.3569759}
\showDOI{\tempurl}


\bibitem[Boersma and Weenink(2021)]%
        {praat}
\bibfield{author}{\bibinfo{person}{Paul Boersma} {and} \bibinfo{person}{David Weenink}.} \bibinfo{year}{2021}\natexlab{}.
\newblock \bibinfo{title}{{P}raat: doing phonetics by computer [{C}omputer program]}.
\newblock \bibinfo{howpublished}{Version 6.1.38, retrieved 2 January 2021 \url{http://www.praat.org/}}.
\newblock


\bibitem[Braun and Clarke(2006)]%
        {thematicAnalysis}
\bibfield{author}{\bibinfo{person}{Virginia Braun} {and} \bibinfo{person}{Victoria Clarke}.} \bibinfo{year}{2006}\natexlab{}.
\newblock \showarticletitle{Using thematic analysis in psychology}.
\newblock \bibinfo{journal}{\emph{Qualitative Research in Psychology}} \bibinfo{volume}{3}, \bibinfo{number}{2} (\bibinfo{year}{2006}), \bibinfo{pages}{77--101}.
\newblock
\urldef\tempurl%
\url{https://doi.org/10.1191/1478088706qp063oa}
\showDOI{\tempurl}
\showeprint{https://www.tandfonline.com/doi/pdf/10.1191/1478088706qp063oa}


\bibitem[Brown et~al\mbox{.}(2020)]%
        {brown2020language}
\bibfield{author}{\bibinfo{person}{Tom~B. Brown}, \bibinfo{person}{Benjamin Mann}, \bibinfo{person}{Nick Ryder}, \bibinfo{person}{Melanie Subbiah}, \bibinfo{person}{Jared Kaplan}, \bibinfo{person}{Prafulla Dhariwal}, \bibinfo{person}{Arvind Neelakantan}, \bibinfo{person}{Pranav Shyam}, \bibinfo{person}{Girish Sastry}, \bibinfo{person}{Amanda Askell}, \bibinfo{person}{Sandhini Agarwal}, \bibinfo{person}{Ariel Herbert-Voss}, \bibinfo{person}{Gretchen Krueger}, \bibinfo{person}{Tom Henighan}, \bibinfo{person}{Rewon Child}, \bibinfo{person}{Aditya Ramesh}, \bibinfo{person}{Daniel~M. Ziegler}, \bibinfo{person}{Jeffrey Wu}, \bibinfo{person}{Clemens Winter}, \bibinfo{person}{Christopher Hesse}, \bibinfo{person}{Mark Chen}, \bibinfo{person}{Eric Sigler}, \bibinfo{person}{Mateusz Litwin}, \bibinfo{person}{Scott Gray}, \bibinfo{person}{Benjamin Chess}, \bibinfo{person}{Jack Clark}, \bibinfo{person}{Christopher Berner}, \bibinfo{person}{Sam McCandlish}, \bibinfo{person}{Alec Radford}, \bibinfo{person}{Ilya Sutskever},
  {and} \bibinfo{person}{Dario Amodei}.} \bibinfo{year}{2020}\natexlab{}.
\newblock \bibinfo{title}{Language Models are Few-Shot Learners}.
\newblock
\newblock
\showeprint[arxiv]{2005.14165}~[cs.CL]


\bibitem[Bubeck et~al\mbox{.}(2023)]%
        {bubeck2023sparks}
\bibfield{author}{\bibinfo{person}{Sébastien Bubeck}, \bibinfo{person}{Varun Chandrasekaran}, \bibinfo{person}{Ronen Eldan}, \bibinfo{person}{Johannes Gehrke}, \bibinfo{person}{Eric Horvitz}, \bibinfo{person}{Ece Kamar}, \bibinfo{person}{Peter Lee}, \bibinfo{person}{Yin~Tat Lee}, \bibinfo{person}{Yuanzhi Li}, \bibinfo{person}{Scott Lundberg}, \bibinfo{person}{Harsha Nori}, \bibinfo{person}{Hamid Palangi}, \bibinfo{person}{Marco~Tulio Ribeiro}, {and} \bibinfo{person}{Yi Zhang}.} \bibinfo{year}{2023}\natexlab{}.
\newblock \bibinfo{title}{Sparks of Artificial General Intelligence: Early experiments with GPT-4}.
\newblock
\newblock
\showeprint[arxiv]{2303.12712}~[cs.CL]


\bibitem[Chung et~al\mbox{.}(2022)]%
        {Chung2022}
\bibfield{author}{\bibinfo{person}{John Joon~Young Chung}, \bibinfo{person}{Wooseok Kim}, \bibinfo{person}{Kang~Min Yoo}, \bibinfo{person}{Hwaran Lee}, \bibinfo{person}{Eytan Adar}, {and} \bibinfo{person}{Minsuk Chang}.} \bibinfo{year}{2022}\natexlab{}.
\newblock \showarticletitle{{TaleBrush: Sketching Stories with Generative Pretrained Language Models}}. In \bibinfo{booktitle}{\emph{Conference on Human Factors in Computing Systems - Proceedings}}. \bibinfo{publisher}{Association for Computing Machinery}, \bibinfo{pages}{19}.
\newblock
\showISBNx{9781450391573}
\urldef\tempurl%
\url{https://doi.org/10.1145/3491102.3501819}
\showDOI{\tempurl}


\bibitem[Comuniqa({[n.\,d.]})]%
        {googleComuniqaApps}
\bibfield{author}{\bibinfo{person}{Comuniqa}.} \bibinfo{year}{[n.\,d.]}\natexlab{}.
\newblock \bibinfo{title}{Comuniqa - Apps on Google Play --- play.google.com}.
\newblock \bibinfo{howpublished}{\url{https://play.google.com/store/apps/details?id=com.comuniqa&hl=en&gl=US}}.
\newblock
\newblock
\shownote{[Accessed 25-01-2024]}.


\bibitem[Giorgino(2009)]%
        {JSSv031i07}
\bibfield{author}{\bibinfo{person}{Toni Giorgino}.} \bibinfo{year}{2009}\natexlab{}.
\newblock \showarticletitle{Computing and Visualizing Dynamic Time Warping Alignments in R: The dtw Package}.
\newblock \bibinfo{journal}{\emph{Journal of Statistical Software}} \bibinfo{volume}{31}, \bibinfo{number}{7} (\bibinfo{year}{2009}).
\newblock
\urldef\tempurl%
\url{https://doi.org/10.18637/jss.v031.i07}
\showDOI{\tempurl}


\bibitem[Goodreads({[n.\,d.]})]%
        {goodreadsPublicSpeaking}
\bibfield{author}{\bibinfo{person}{Inc. Goodreads}.} \bibinfo{year}{[n.\,d.]}\natexlab{}.
\newblock \bibinfo{title}{{T}he {A}rt of {P}ublic {S}peaking --- goodreads.com}.
\newblock \bibinfo{howpublished}{\url{https://www.goodreads.com/en/book/show/3363618}}.
\newblock
\newblock
\shownote{[Accessed 26-01-2024]}.


\bibitem[Google({[n.\,d.]})]%
        {googleAndroidApps}
\bibfield{author}{\bibinfo{person}{Google}.} \bibinfo{year}{[n.\,d.]}\natexlab{}.
\newblock \bibinfo{title}{Android Apps on Google Play --- play.google.com}.
\newblock \bibinfo{howpublished}{\url{https://play.google.com/store/}}.
\newblock
\newblock
\shownote{[Accessed 25-01-2024]}.


\bibitem[Jadoul et~al\mbox{.}(2018)]%
        {parselmouth}
\bibfield{author}{\bibinfo{person}{Yannick Jadoul}, \bibinfo{person}{Bill Thompson}, {and} \bibinfo{person}{Bart de Boer}.} \bibinfo{year}{2018}\natexlab{}.
\newblock \showarticletitle{Introducing {P}arselmouth: A {P}ython interface to {P}raat}.
\newblock \bibinfo{journal}{\emph{Journal of Phonetics}}  \bibinfo{volume}{71} (\bibinfo{year}{2018}), \bibinfo{pages}{1--15}.
\newblock
\urldef\tempurl%
\url{https://doi.org/10.1016/j.wocn.2018.07.001}
\showDOI{\tempurl}


\bibitem[Jakesch et~al\mbox{.}(2023)]%
        {Jakesch2023}
\bibfield{author}{\bibinfo{person}{Maurice Jakesch}, \bibinfo{person}{Advait Bhat}, \bibinfo{person}{Daniel Buschek}, \bibinfo{person}{Lior Zalmanson}, {and} \bibinfo{person}{Mor Naaman}.} \bibinfo{year}{2023}\natexlab{}.
\newblock \showarticletitle{{Co-Writing with Opinionated Language Models Affects Users' Views}}. In \bibinfo{booktitle}{\emph{Conference on Human Factors in Computing Systems - Proceedings}}. \bibinfo{publisher}{Association for Computing Machinery}, \bibinfo{pages}{15}.
\newblock
\showISBNx{9781450394215}
\urldef\tempurl%
\url{https://doi.org/10.1145/3544548.3581196}
\showDOI{\tempurl}
\showeprint[arxiv]{2302.00560}


\bibitem[Junaidi(2020)]%
        {junaidi2020artificial}
\bibfield{author}{\bibinfo{person}{Junaidi Junaidi}.} \bibinfo{year}{2020}\natexlab{}.
\newblock \showarticletitle{Artificial intelligence in EFL context: rising students’ speaking performance with Lyra virtual assistance}.
\newblock \bibinfo{journal}{\emph{International Journal of Advanced Science and Technology Rehabilitation}} \bibinfo{volume}{29}, \bibinfo{number}{5} (\bibinfo{year}{2020}), \bibinfo{pages}{6735--6741}.
\newblock


\bibitem[Li et~al\mbox{.}(2023)]%
        {li2023collaborative}
\bibfield{author}{\bibinfo{person}{Qintong Li}, \bibinfo{person}{Leyang Cui}, \bibinfo{person}{Lingpeng Kong}, {and} \bibinfo{person}{Wei Bi}.} \bibinfo{year}{2023}\natexlab{}.
\newblock \bibinfo{title}{Collaborative Evaluation: Exploring the Synergy of Large Language Models and Humans for Open-ended Generation Evaluation}.
\newblock
\newblock
\showeprint[arxiv]{2310.19740}~[cs.CL]


\bibitem[Louradour(2023)]%
        {lintoai2023whispertimestamped}
\bibfield{author}{\bibinfo{person}{J{\'e}r{\^o}me Louradour}.} \bibinfo{year}{2023}\natexlab{}.
\newblock \bibinfo{title}{whisper-timestamped}.
\newblock \bibinfo{howpublished}{\url{https://github.com/linto-ai/whisper-timestamped}}.
\newblock


\bibitem[LTD({[n.\,d.]})]%
        {EnglishYaari}
\bibfield{author}{\bibinfo{person}{APNI YAARI EDUCATION~PVT. LTD}.} \bibinfo{year}{[n.\,d.]}\natexlab{}.
\newblock \bibinfo{howpublished}{\url{https://www.englishyaari.com/}}.
\newblock


\bibitem[Ltd({[n.\,d.]})]%
        {IELTSSpeaking}
\bibfield{author}{\bibinfo{person}{IDP~Education Ltd}.} \bibinfo{year}{[n.\,d.]}\natexlab{}.
\newblock \bibinfo{title}{ielts-band-scores}.
\newblock \bibinfo{howpublished}{\url{https://ieltsjp.com/japan/about/about-ielts/ielts-band-scores/en-gb}}.
\newblock


\bibitem[Microsoft({[n.\,d.]})]%
        {MicrosoftFillerWords}
\bibfield{author}{\bibinfo{person}{Microsoft}.} \bibinfo{year}{[n.\,d.]}\natexlab{}.
\newblock \bibinfo{title}{Recognizing and avoiding filler words}.
\newblock \bibinfo{howpublished}{\url{https://www.microsoft.com/en-us/microsoft-365-life-hacks/writing/recognizing-avoiding-filler-words}}.
\newblock


\bibitem[Montgomerie({[n.\,d.]})]%
        {cefrPredictor}
\bibfield{author}{\bibinfo{person}{Adam Montgomerie}.} \bibinfo{year}{[n.\,d.]}\natexlab{}.
\newblock \bibinfo{howpublished}{\url{https://amontgomerie.github.io/2021/03/14/cefr-level-prediction.html}}.
\newblock


\bibitem[Nass and Moon(2000)]%
        {machineAndMindlessness}
\bibfield{author}{\bibinfo{person}{Clifford Nass} {and} \bibinfo{person}{Youngme Moon}.} \bibinfo{year}{2000}\natexlab{}.
\newblock \showarticletitle{Machines and Mindlessness: Social Responses to Computers}.
\newblock \bibinfo{journal}{\emph{The Society for the Psychological Study of Social Issues}}  \bibinfo{volume}{56} (\bibinfo{year}{2000}).
\newblock
Issue 1.
\urldef\tempurl%
\url{https://doi.org/10.1111/0022-4537.00153}
\showDOI{\tempurl}


\bibitem[Pearson({[n.\,d.]})]%
        {cefrWordLevel}
\bibfield{author}{\bibinfo{person}{Pearson}.} \bibinfo{year}{[n.\,d.]}\natexlab{}.
\newblock \bibinfo{howpublished}{\url{https://www.english.com/gse/teacher-toolkit/user/vocabulary}}.
\newblock


\bibitem[Peng et~al\mbox{.}(2023)]%
        {peng2023storyfier}
\bibfield{author}{\bibinfo{person}{Zhenhui Peng}, \bibinfo{person}{Xingbo Wang}, \bibinfo{person}{Qiushi Han}, \bibinfo{person}{Junkai Zhu}, \bibinfo{person}{Xiaojuan Ma}, {and} \bibinfo{person}{Huamin Qu}.} \bibinfo{year}{2023}\natexlab{}.
\newblock \showarticletitle{Storyfier: Exploring Vocabulary Learning Support with Text Generation Models}. In \bibinfo{booktitle}{\emph{Proceedings of the 36th Annual ACM Symposium on User Interface Software and Technology}} (<conf-loc>, <city>San Francisco</city>, <state>CA</state>, <country>USA</country>, </conf-loc>) \emph{(\bibinfo{series}{UIST '23})}. \bibinfo{publisher}{Association for Computing Machinery}, \bibinfo{address}{New York, NY, USA}, Article \bibinfo{articleno}{46}, \bibinfo{numpages}{16}~pages.
\newblock
\showISBNx{9798400701320}
\urldef\tempurl%
\url{https://doi.org/10.1145/3586183.3606786}
\showDOI{\tempurl}


\bibitem[Petridis et~al\mbox{.}(2023)]%
        {Petridis2023}
\bibfield{author}{\bibinfo{person}{Savvas Petridis}, \bibinfo{person}{Nicholas Diakopoulos}, \bibinfo{person}{Kevin Crowston}, \bibinfo{person}{Mark Hansen}, \bibinfo{person}{Keren Henderson}, \bibinfo{person}{Stan Jastrzebski}, \bibinfo{person}{Jeffrey~V. Nickerson}, {and} \bibinfo{person}{Lydia~B. Chilton}.} \bibinfo{year}{2023}\natexlab{}.
\newblock \showarticletitle{{AngleKindling: Supporting Journalistic Angle Ideation with Large Language Models}}. In \bibinfo{booktitle}{\emph{Conference on Human Factors in Computing Systems - Proceedings}}. \bibinfo{publisher}{Association for Computing Machinery}, \bibinfo{pages}{16}.
\newblock
\showISBNx{9781450394215}
\urldef\tempurl%
\url{https://doi.org/10.1145/3544548.3580907}
\showDOI{\tempurl}


\bibitem[Radford et~al\mbox{.}(2023)]%
        {radford2023whisper}
\bibfield{author}{\bibinfo{person}{Alec Radford}, \bibinfo{person}{Jong~Wook Kim}, \bibinfo{person}{Tao Xu}, \bibinfo{person}{Greg Brockman}, \bibinfo{person}{Christine McLeavey}, {and} \bibinfo{person}{Ilya Sutskever}.} \bibinfo{year}{2023}\natexlab{}.
\newblock \showarticletitle{Robust speech recognition via large-scale weak supervision}. In \bibinfo{booktitle}{\emph{Proceedings of the 40th International Conference on Machine Learning}} (Honolulu, Hawaii, USA) \emph{(\bibinfo{series}{ICML'23})}. \bibinfo{publisher}{JMLR.org}, Article \bibinfo{articleno}{1182}, \bibinfo{numpages}{27}~pages.
\newblock


\bibitem[Ruan et~al\mbox{.}(2019)]%
        {Ruan2019}
\bibfield{author}{\bibinfo{person}{Sherry Ruan}, \bibinfo{person}{Liwei Jiang}, \bibinfo{person}{Justin Xu}, \bibinfo{person}{Bryce Joe~Kun Tham}, \bibinfo{person}{Zhengneng Qiu}, \bibinfo{person}{Yeshuang Zhu}, \bibinfo{person}{Elizabeth~L. Murnane}, \bibinfo{person}{Emma Brunskill}, {and} \bibinfo{person}{James~A. Landay}.} \bibinfo{year}{2019}\natexlab{}.
\newblock \showarticletitle{{QuizBot: A Dialogue-based Adaptive Learning System for Factual Knowledge}}. In \bibinfo{booktitle}{\emph{Conference on Human Factors in Computing Systems - Proceedings}}, Vol.~\bibinfo{volume}{13}. \bibinfo{publisher}{Association for Computing Machinery}.
\newblock
\showISBNx{9781450359702}
\urldef\tempurl%
\url{https://doi.org/10.1145/3290605.3300587}
\showDOI{\tempurl}


\bibitem[Ruan et~al\mbox{.}(2021)]%
        {ruan2021englishbot}
\bibfield{author}{\bibinfo{person}{Sherry Ruan}, \bibinfo{person}{Liwei Jiang}, \bibinfo{person}{Qianyao Xu}, \bibinfo{person}{Zhiyuan Liu}, \bibinfo{person}{Glenn~M Davis}, \bibinfo{person}{Emma Brunskill}, {and} \bibinfo{person}{James~A. Landay}.} \bibinfo{year}{2021}\natexlab{}.
\newblock \showarticletitle{EnglishBot: An AI-Powered Conversational System for Second Language Learning}. In \bibinfo{booktitle}{\emph{26th International Conference on Intelligent User Interfaces}} (College Station, TX, USA) \emph{(\bibinfo{series}{IUI '21})}. \bibinfo{publisher}{Association for Computing Machinery}, \bibinfo{address}{New York, NY, USA}, \bibinfo{pages}{434–444}.
\newblock
\showISBNx{9781450380171}
\urldef\tempurl%
\url{https://doi.org/10.1145/3397481.3450648}
\showDOI{\tempurl}


\bibitem[Shen and Wu(2023)]%
        {shen2023parachute}
\bibfield{author}{\bibinfo{person}{Hua Shen} {and} \bibinfo{person}{Tongshuang Wu}.} \bibinfo{year}{2023}\natexlab{}.
\newblock \bibinfo{title}{Parachute: Evaluating Interactive Human-LM Co-writing Systems}.
\newblock
\newblock
\showeprint[arxiv]{2303.06333}~[cs.HC]


\bibitem[Speech({[n.\,d.]})]%
        {azureSpeech}
\bibfield{author}{\bibinfo{person}{Azure Speech}.} \bibinfo{year}{[n.\,d.]}\natexlab{}.
\newblock \bibinfo{howpublished}{\url{https://azure.microsoft.com/en-us/products/ai-services/ai-speech/}}.
\newblock


\bibitem[Tanveer et~al\mbox{.}(2015)]%
        {tanveer2015rhema}
\bibfield{author}{\bibinfo{person}{M.~Iftekhar Tanveer}, \bibinfo{person}{Emy Lin}, {and} \bibinfo{person}{Mohammed~(Ehsan) Hoque}.} \bibinfo{year}{2015}\natexlab{}.
\newblock \showarticletitle{Rhema: A Real-Time In-Situ Intelligent Interface to Help People with Public Speaking}. In \bibinfo{booktitle}{\emph{Proceedings of the 20th International Conference on Intelligent User Interfaces}} (Atlanta, Georgia, USA) \emph{(\bibinfo{series}{IUI '15})}. \bibinfo{publisher}{Association for Computing Machinery}, \bibinfo{address}{New York, NY, USA}, \bibinfo{pages}{286–295}.
\newblock
\showISBNx{9781450333061}
\urldef\tempurl%
\url{https://doi.org/10.1145/2678025.2701386}
\showDOI{\tempurl}


\bibitem[Trinh et~al\mbox{.}(2017)]%
        {Trinh2017robocop}
\bibfield{author}{\bibinfo{person}{H. Trinh}, \bibinfo{person}{R. Asadi}, \bibinfo{person}{D. Edge}, {and} \bibinfo{person}{T. Bickmore}.} \bibinfo{year}{2017}\natexlab{}.
\newblock \showarticletitle{RoboCOP: A Robotic Coach for Oral Presentations}.
\newblock \bibinfo{journal}{\emph{Proc. ACM Interact. Mob. Wearable Ubiquitous Technol.}} \bibinfo{volume}{1}, \bibinfo{number}{2}, Article \bibinfo{articleno}{27} (\bibinfo{date}{jun} \bibinfo{year}{2017}), \bibinfo{numpages}{24}~pages.
\newblock
\urldef\tempurl%
\url{https://doi.org/10.1145/3090092}
\showDOI{\tempurl}


\bibitem[Tuhovsky(2015)]%
        {tuhovsky2015communication}
\bibfield{author}{\bibinfo{person}{I. Tuhovsky}.} \bibinfo{year}{2015}\natexlab{}.
\newblock \bibinfo{booktitle}{\emph{Communication Skills: A Practical Guide to Improving Your Social Intelligence, Presentation, Persuasion and Public Speaking}}.
\newblock \bibinfo{publisher}{CreateSpace Independent Publishing Platform}.
\newblock
\showISBNx{9781515031918}
\urldef\tempurl%
\url{https://books.google.co.in/books?id=WBsGswEACAAJ}
\showURL{%
\tempurl}


\bibitem[Valencia et~al\mbox{.}(2023)]%
        {Valencia2023}
\bibfield{author}{\bibinfo{person}{Stephanie Valencia}, \bibinfo{person}{Richard Cave}, \bibinfo{person}{Krystal Kallarackal}, \bibinfo{person}{Katie Seaver}, \bibinfo{person}{Michael Terry}, {and} \bibinfo{person}{Shaun~K. Kane}.} \bibinfo{year}{2023}\natexlab{}.
\newblock \showarticletitle{{"The less I type, the better": How AI Language Models can Enhance or Impede Communication for AAC Users}}. In \bibinfo{booktitle}{\emph{Conference on Human Factors in Computing Systems - Proceedings}}. \bibinfo{publisher}{Association for Computing Machinery}, \bibinfo{pages}{14}.
\newblock
\showISBNx{9781450394215}
\urldef\tempurl%
\url{https://doi.org/10.1145/3544548.3581560}
\showDOI{\tempurl}


\bibitem[Wang et~al\mbox{.}(2020)]%
        {wang2020voicecoach}
\bibfield{author}{\bibinfo{person}{Xingbo Wang}, \bibinfo{person}{Haipeng Zeng}, \bibinfo{person}{Yong Wang}, \bibinfo{person}{Aoyu Wu}, \bibinfo{person}{Zhida Sun}, \bibinfo{person}{Xiaojuan Ma}, {and} \bibinfo{person}{Huamin Qu}.} \bibinfo{year}{2020}\natexlab{}.
\newblock \showarticletitle{VoiceCoach: Interactive Evidence-based Training for Voice Modulation Skills in Public Speaking}. In \bibinfo{booktitle}{\emph{Proceedings of the 2020 CHI Conference on Human Factors in Computing Systems}} (, Honolulu, HI, USA,) \emph{(\bibinfo{series}{CHI '20})}. \bibinfo{publisher}{Association for Computing Machinery}, \bibinfo{address}{New York, NY, USA}, \bibinfo{pages}{1–12}.
\newblock
\showISBNx{9781450367080}
\urldef\tempurl%
\url{https://doi.org/10.1145/3313831.3376726}
\showDOI{\tempurl}


\bibitem[Wei et~al\mbox{.}(2022)]%
        {wei2022emergent}
\bibfield{author}{\bibinfo{person}{Jason Wei}, \bibinfo{person}{Yi Tay}, \bibinfo{person}{Rishi Bommasani}, \bibinfo{person}{Colin Raffel}, \bibinfo{person}{Barret Zoph}, \bibinfo{person}{Sebastian Borgeaud}, \bibinfo{person}{Dani Yogatama}, \bibinfo{person}{Maarten Bosma}, \bibinfo{person}{Denny Zhou}, \bibinfo{person}{Donald Metzler}, \bibinfo{person}{Ed~H. Chi}, \bibinfo{person}{Tatsunori Hashimoto}, \bibinfo{person}{Oriol Vinyals}, \bibinfo{person}{Percy Liang}, \bibinfo{person}{Jeff Dean}, {and} \bibinfo{person}{William Fedus}.} \bibinfo{year}{2022}\natexlab{}.
\newblock \bibinfo{title}{Emergent Abilities of Large Language Models}.
\newblock
\newblock
\showeprint[arxiv]{2206.07682}~[cs.CL]


\bibitem[Yuan et~al\mbox{.}(2022)]%
        {yuan2022wordcraft}
\bibfield{author}{\bibinfo{person}{Ann Yuan}, \bibinfo{person}{Andy Coenen}, \bibinfo{person}{Emily Reif}, {and} \bibinfo{person}{Daphne Ippolito}.} \bibinfo{year}{2022}\natexlab{}.
\newblock \showarticletitle{Wordcraft: Story Writing With Large Language Models}. In \bibinfo{booktitle}{\emph{27th International Conference on Intelligent User Interfaces}} (Helsinki, Finland) \emph{(\bibinfo{series}{IUI '22})}. \bibinfo{publisher}{Association for Computing Machinery}, \bibinfo{address}{New York, NY, USA}, \bibinfo{pages}{841–852}.
\newblock
\showISBNx{9781450391443}
\urldef\tempurl%
\url{https://doi.org/10.1145/3490099.3511105}
\showDOI{\tempurl}


\end{thebibliography}


\appendix
\section{Questionnaires}\label{sec:appendix_questions}
\subsection{Pre-Interview Questions}
\begin{enumerate}
    \item When did you start learning and practicing English? Could you share your preferred methods for practicing English?
    \item Have you received feedback on your English language skills from experts or peers?
    \item Can you mention any specific technology or tools you use to enhance your communication and speaking skills?
    \item Tools or Apps used?
        \begin{enumerate}
            \item Do you engage in self-recording to improve your communication skills?
            \item Do you use ChatGPT or other AI based tools for verbal or written skills?
            \item Are these tools or apps free or paid?
        \end{enumerate}
    \item When it comes to language learning and communication, do you prefer using mobile apps or web-based platforms?
    \item Would like to Share experience with respect to the above application you have used
    \item Have you faced any challenges or gaps in your communication or speaking skills?
    \item Is there a particular aspect, verbal or non-verbal, that you find more challenging?
    \item Are you focused on improving spoken or written communication, or both?
    \item What motivates you to improve these skills?
    \item Do you Prefer human or AI? Why
\end{enumerate}

\subsection{Post-Interview Questions for only ComuniQa.ai based Group}
\begin{enumerate}
    \item Can you briefly describe your experience with ComuniQa.ai in improving your speaking and communication skills
    \item In your opinion, what were the key benefits of using ComuniQa.ai for improving your speaking and communication skills?
    \item Were there any challenges or limitations you encountered while using ComuniQa.ai? Please describe.
    \begin{enumerate}
        \item Pros and Cons
    \end{enumerate}
    \item Was the feedback actionable given by the app ?
    \item What are your thoughts on the comparative effectiveness of live feedback and end-of-session feedback for improving speaking skills? Please share your insights on the advantages and disadvantages of each approach. (old: Do you think live feedback will be better than feedback at the end?)
    \item Where would you like to use such an AI tool?
    \item How frequently have you used AI/LLM tools for this purpose? (Likert Scale)
    \item How did you find the overall study experience, including the learning sessions with human experts and ComuniQa.ai tools?
    \item Were the benefits offered in exchange for participation in the study (e.g., gift vouchers, career guidance) valuable to you?
    \item Is there anything else you would like to share, or any additional comments or feedback related to the study or ComuniQa.ai app for speaking communication skills?    
\end{enumerate}

\subsection{Post-Interview questions for only Human Expert Group}
\begin{enumerate}
    \item Can you briefly describe your experience with human expert guidance in improving your speaking and communication skills?
    \item In your opinion, what were the key benefits of using human experts for enhancing your speaking and communication skills?
    \item Was the interaction with the human expert useful in improving your spoken skills? How?
    \item Were there any challenges or limitations you encountered with human experts for this purpose? Please describe.
    \item Was the feedback actionable provided by the human experts?
    \item Pros and Cons:
        \begin{enumerate}
            \item Pros: What were the positive aspects of receiving guidance exclusively from human experts?
            \item Cons: Were there any drawbacks or limitations in the feedback provided by human experts?
        \end{enumerate}
    \item How frequently did you exclusively rely on human experts for this purpose? (Likert Scale: 1-5)
    \item Were the benefits offered in exchange for participation in the study (e.g., gift vouchers, career guidance) valuable to you?
    \item Is there anything else you would like to share, or any additional comments or feedback related to the study or the use of human experts for improving speaking communication skills?
\end{enumerate}

\subsection{Post-Interview questions for Human Expert \& ComuniQa.ai based Group}
\begin{enumerate}
    \item Can you briefly describe your experience with AI tools in improving your speaking and communication skills
    \item In your opinion, what were the key benefits of using ComuniQa.ai tool for improving your speaking and communication skills?
    \item Was the interaction with Expert useful in improving spoken skills? How?
    \item Were there any challenges or limitations you encountered while using  ComuniQa.ai tool for this purpose? Please describe.
    \begin{enumerate}
        \item Pros and Cons
    \end{enumerate}
    \item Was the feedback actionable?
    \item Expert Feedback
    \begin{enumerate}
        \item pros and cons
        \item Was the feedback actionable?
    \end{enumerate}
    \item Can you compare the feedback and guidance provided by AI/LLM tools to that of human experts? Are there any differences in the quality or effectiveness of feedback?
    \item What are your thoughts on the comparative effectiveness of live feedback and end-of-session feedback for improving speaking skills? Please share your insights on the advantages and disadvantages of each approach. (old: Do you think live feedback will be better than feedback at the end?)
    \item What are your preferences when it comes to receiving feedback on your speaking and communication skills: AI/LLM or human instructors? Why?
    \item What are your preferences when it comes to receiving feedback on your speaking and communication skills: AI/LLM or human instructors? Why?
    \item Where would you like to use such an AI tool?
    \item How frequently have you used AI/LLM tools for this purpose? (Likert Scale)
    \item How did you find the overall study experience, including the learning sessions with human experts and AI/LLM tools?
    \item Were the benefits offered in exchange for participation in the study (e.g., gift vouchers, career guidance) valuable to you?
    \item Is there anything else you would like to share, or any additional comments or feedback related to the study or AI/LLM tools for speaking communication skills?
\end{enumerate}

\newpage
\onecolumn
\section{Participant Demographics}\label{sec:appendix_demographics}
\begin{table}[ht]
    \centering
    \scriptsize
    \begin{tabular}{|p{2cm}|p{3cm}|l|l|l|p{3cm}|l|}
    \hline
        \textbf{Name} & \textbf{Program} & \textbf{Native Language} & \textbf{Age} & \textbf{Gender} & \textbf{Motivation} & \textbf{Proficiency} \\ 
        \hline
        P1 & Mtech – First year & Malayalam & 22 & Male & Public Speaking or Class Presentations & 4 \\ \hline
        P2 & Btech - Fourth Year & Hindi & 21 & Female & Public Speaking or Class Presentations & 3 \\ \hline
        P3 & Mtech – Second year & Gujrati & 23 & Male & Daily Conversations & 2 \\ \hline
        P4 & Btech - Third Year & Marathi & 21 & Male & Interview Practice & 3 \\ \hline
        P5 & Btech - Third Year & Hindi & 21 & Male & Public Speaking or Class Presentations & 3 \\ \hline
        P6 & Btech - Fourth Year & Malayalam & 21 & Female & Preparation for Tests (IELTS/TOEFL) & 4 \\ \hline
        P7 & Btech - Third Year & Hindi & 20 & Male & Public Speaking or Class Presentations & 2 \\ \hline
        P8 & Btech - Third Year & Hindi & 21 & Male & Interview Practice & 3 \\ \hline
        P9 & Btech - Third Year & Hindi & 22 & Male & Interview Practice & 3 \\ \hline
        P10 & Btech - Third Year & Hindi & 20 & Male & Interview Practice & 4 \\ \hline
        P11 & Btech - Third Year & Hindi & 21 & Male & Public Speaking or Class Presentations & ~ \\ \hline
        P12 & Btech - Third Year & English & 20 & Female & Interview Practice & 4 \\ \hline
        P13 & Btech - Second Year & Hindi & 20 & Male & Preparation for Tests (IELTS/TOEFL) & 4 \\ \hline
        P14 & Btech - First Year & Hindi & 18 & Male & Public Speaking or Class Presentations & 3 \\ \hline
        P15 & Btech - First Year & Hindi & 18 & Male & Public Speaking or Class Presentations & 2 \\ \hline
        P16 & Mtech – Second year & Telugu & 23 & Male & Daily Conversations & 4 \\ \hline
        P17 & Btech - First Year & Hindi & 18 & Male & Interview Practice & 1 \\ \hline
        P18 & Btech - First Year & Hindi & 18 & Male & Interview practice and Public speaking. & 3 \\ \hline
        P19 & Btech - First Year & English & 18 & Male & Interview Practice & 3 \\ \hline
        P20 & Btech - First Year & Hindi & 18 & Male & Public Speaking or Class Presentations & 3 \\ \hline
        P21 & Btech - First Year & Hindi & 19 & Male & Daily Conversations & 3 \\ \hline
        P22 & Btech - First Year & Hindi & 18 & Male & Daily Conversations & 3 \\ \hline
        P23 & Btech - First Year & Hindi & 18 & Male & Daily Conversations & 2 \\ \hline
        P24 & Btech - First Year & Bengali & 18 & Male & Interview Practice & 4 \\ \hline
        P25 & Btech - Second Year & English & 19 & Female & Public Speaking or Class Presentations & 4 \\ \hline
        P26 & Btech - First Year & Hindi & 19 & Male & Public Speaking or Class Presentations & 4 \\ \hline
        P27 & Btech - First Year & Hindi & 18 & Male & Daily Conversations & 3 \\ \hline
        P28 & Btech - Third Year & Hindi & 20 & Male & Public Speaking or Class Presentations & 3 \\ \hline
        P29 & Btech - Fourth Year & Hindi & 24 & Male & Interview Practice & 4 \\ \hline
        P30 & Btech - Fourth Year & Hindi & 22 & Female & Public Speaking or Class Presentations & 4 \\ \hline
        P31 & Btech - First Year & Hindi & 19 & Male & Interview Practice & 2 \\ \hline
        P32 & Btech - Second Year & Hindi & 18 & Male & None & 1 \\ \hline
        P33 & Btech - Second Year & Punjabi & 19 & Male & Preparation for Tests (IELTS/TOEFL) & 3 \\ \hline
        P34 & Btech - Third Year & Hindi & 20 & Female & Daily Conversations & 3 \\ \hline
    \end{tabular}
    \caption{Participant Demographics}
    \label{table:participant_demographics}
\end{table}

\end{document}